\documentstyle[preprint,pra,aps,eqsecnum]{revtex}
\tightenlines
\begin{document}
%\tightenlines
\draft

\title{Quantum Theory of a High Harmonic Generation
\\
as a Three-Step Process}

\author{M.~Yu.~Kuchiev and V.~N.~Ostrovsky\cite{SP}}

\address{School of Physics, University of New South Wales
Sydney 2052, Australia}

\maketitle

\begin{abstract}
Fully quantum treatment explicitly presents 
the high harmonic generation as a three-step process: 
(i) above threshold ionization (ATI) is followed by (ii) 
electron propagation in a laser-dressed continuum. Subsequently 
(iii) stimulated (or laser assisted) recombination brings
the electron back into the initial state with emission of
a high-energy photon. Contributions
of all ATI channels add up coherently. All three stages of
the process are described by simple, mostly analytical
expressions that allow a detailed physical interpretation. 
A very good quantitative agreement with 
the previous calculations on the harmonic generation by 
H$^-$ ion is demonstrated, thus supplementing the conceptual 
significance of the theory with its practical efficiency.
The virtue of the present scheme is further supported 
by a good accord between the calculations in length 
and velocity gauges for the high-energy photon.
\end{abstract}


%\twocolumn
%\narrowtext

\section{INTRODUCTION} \label{Int}

Under the influence of an intensive electromagnetic field an atom 
can emit electrons and photons. 
The number of photons absorbed from the field in the first process
generally can exceed the minimum necessary for ionization
resulting in distribution of the photoelectrons over 
the above threshold ionization (ATI) channels.
The photon production manifests itself as the harmonics 
generation (HG) for the incident monochromatic laser radiation. 
Both ATI and HG are capable of populating the channels with 
remarkably high energy, as has recently been registered in 
experiments (see, e.g., Ref.\cite{eh,Mac,er,temporal,Zhou}) 
and tackled by the theory 
\cite{Eb,KulShore,C,KulanderA,KulanderB,Hu,Lew,Bothers,Brapid,B,Zar,PlatStr,ellip,Prot,Buni,SandnerA,SandnerB,Beckertwocenter}
(the list of references is unavoidably incomplete, for more 
bibliography see the reviews \cite{Protrev,Strelkov}).

An idea that the two processes referred above are interrelated 
was articulated long ago. Since in the HG process
an active electron ends up in the initial bound state,
it is appealing to represent it as ionization followed by
recombination. This mechanism presumes a strong interaction
between the emitted electron and the core that is omitted
in the standard Keldysh \cite{Keldysh} model of multiphoton
ionization. The importance of this interaction 
was first pointed out by Kuchiev \cite{Ku87}, 
who predicted several phenomena for which the electron-core 
interaction plays a crucial role. The related 
mechanism was named {\it ``atomic antenna''}.

Specifically for HG, the simple relation between this process 
and ATI was suggested by Eberly {\it et al}\/ \cite{Eb} but 
proved to be non-realistic, see below. 
The hybrid classical-quantum model due to Corkum \cite{C} 
(see also the paper by Kulander {\it et al}\/ \cite{KulanderB}) 
casts HG as a three-step process: tunneling ionization
and subsequent propagation in the continuum is completed by 
recombination. This intuitive model has influenced many
research in experiment and theory. The simplicity
of the model is due to some drastic presumptions.
Usually it is emphasized that the intermediate electron 
propagation in the laser field is described by the 
Corkum \cite{C} model classically. Probably less
attention is paid to the fact that neither the tunneling 
ionization through the time-dependent barrier, nor 
the laser-stimulated recombination receive a genuine 
quantum treatment as well. Being successfully applied to 
the comparison with some experimental data, the 
model resorts to such a loosely defined free parameter as the 
'transverse spread of the electron wave function'. From 
the conceptual side the Corkum \cite{C} model
does not appeal to ATI process just because the
discrete ATI channels do not appear within the classical framework. 
The subsequent theoretical developments were based on more 
sophisticated approaches and led to important advancements 
\cite{Hu,Lew,Bothers,B},
but apparently abandoned a perspective to establish a quantitative 
relation between ATI and HG. The ATI characteristics
merely do not emerge in the papers devoted to HG theory,
%!   below
%!   pervaya iz nizheprivedennykh ssylok vklyuchena neosnovatel'no,
%!   poskol'ku u menya net etoi stat'i 
%!
with few exceptions \cite{Brapid,Zar}.
For instance, Ref.\cite{Buni} establishes what the authors
think to be 'the most general formal relation between
ionization and high harmonic generation' that contains
functional derivative of 'the ground state persistence
amplitude' $Z$ but not ATI amplitude. As the authors
recognize, this relation is 'of limited practical use'
since it requires knowledge of $Z$ for {\it arbitrary}\/
electric field ${\bf F}(t)$.
%!    above 

Theoretically HG is governed by the Fourier components
in the field-dresses wave function.
The proper description of the high-order components
is a challenging task for the theory.
Although some important features could be understood in
classical calculations, the significance of fully quantum 
quantitative theory could not be overestimated.
The computer-intensive numerical studies achieved
substantial progress \cite{KulShore,KulanderA}, 
but the available computer facilities 
often limit them to the one-dimensional models
\cite{Eb,SandnerA,SandnerB}.  
They could be favorably complemented by analytical studies 
capable of providing an important physical insight.
However, with the two different analytic quantum approaches 
being developed recently \cite{Lew,B},
the problem could not yet be considered as closed.
For instance, as far as we know, these two approaches have
never been applied to the same system in order to provide
quantitative comparison of their results, albeit the differences 
in the formulation were indicated by Becker {\it et al}\/ \cite{B}.
%!    below
The only exception known to us is provided by Ref.~\cite{Buni}
where in Fig.~6 some comparison is presented, albeit not for
such a basic characteristics as HG rates, but for harmonic
ellipticity and offset angle that reflect rather subtle
peculiarities of HG process. For the offset angle a severe
disagreement between the two theories is demonstrated.
The lack of comparison looks quite typical to the current
%!   above
state of theory when the attention is mostly directed
towards the qualitative aspects of the problem such 
as description of the pattern of emitted harmonic spectrum.

The principal objective of the present study is to derive a fully 
quantum formulation for the HG amplitude in terms of the ATI amplitude 
and the amplitude of electron laser assisted recombination (LAR) 
in the laser field. Importantly, all the amplitudes are physical, 
i.e. no off-energy-shell entities appear. This circumstance adds to 
the conceptual appeal of the present theory its significance
as a true working tool. In our approach, briefly described
in the Letter to Editor \cite{Lett}, we base on the ideas
outlined \cite{Ku87} and later developed in more detail 
\cite{K95,K96} by Kuchiev.
We successfully test its efficiency by 
comparison with the benchmark calculations
by Becker {\it et al}\/ \cite{B} for HG by H$^-$ ion. 
It should be emphasized that we test and achieve
quantitative agreement for absolute values of HG rates, 
in contrast to many theoretical
works which concentrate mostly on the qualitative issues.
The quantitative character of our development is further
illustrated by comparative calculations within the 
dipole-length and dipole-velocity gauges for the emitted 
high-energy photons: the good agreement testifies in favor
of the method.
In the broader perspective it should be emphasized that our
theoretical technique is directly applicable 
to other processes of current key interest, such as 
multiple ionization by laser radiation or enhanced
population of high ATI channels due to the photoelectron 
rescattering on the atomic core. 

%! below
We start (Sec.\ref{Sec2}) with exposure of the general relations 
and outline our basic approximations. In Section \ref{Sec3}
we cast the harmonic generation amplitude as a sum over contributions 
coming from different ATI channels. The form and interpretation 
of this representation becomes particularly transparent 
when we resort to the adiabatic approximation (Sec.\ref{Secad}). 
The general theory is illustrated by the quantitative results 
in Section \ref{Sec5} that is followed by the concluding discussion 
(Sec.\ref{Sec6}).
%! above

\section{FORMULATION OF THE PROBLEM AND BASIC 
APPROXIMATIONS} \label{Sec2}

\subsection{General relations}

The generation of a harmonic with the frequency $\Omega$ is governed by 
the Fourier transform of the dipole transition matrix element
\begin{eqnarray} \label{Mdef}
D(\Omega) & = & \int^\infty_{-\infty} dt \, \exp ( i \Omega t) \, d(t) ~, 
\\ \label{dtdef}
d(t) & = &
\int d^3 {\bf r} \, \Psi_f ({\bf r}, \, t)^* \, 
\hat{d}_{\mbox {\boldmath $\epsilon$}} \, 
\Psi_i ({\bf r}, \, t) ~,
\hspace{5mm}
\hat{d}_{\mbox {\boldmath $\epsilon$}} = 
{\mbox {\boldmath $\epsilon$}} \cdot {\bf r} ~,
\end{eqnarray} 
where
%$e = -1$ for the electron,
$\Psi_i$ and $\Psi_f$ are initial and final states of the
atomic system dressed by the laser field (atomic units
are used throughout the paper).

We construct the initial field-dressed state $\Psi_i$   
\begin{eqnarray} \label{wfg}
\Psi_i( {\bf r}, t ) & = & 
\Phi_a({\bf r}, t) + \int_{- \infty}^t dt^\prime \:
\int d^3 {\bf r}^\prime \, 
G( {\bf r}, t ; \, {\bf r}^\prime, t^\prime) \,
V_F({\bf r}^\prime, t^\prime) \,  
\Phi_a({\bf r}^\prime, t^\prime) ~,
\end{eqnarray} 
developed out the initial field-free stationary state $\Phi_a$
\begin{eqnarray}\label{Psia}
\Phi_a({\bf r}, t) & = & \varphi_a({\bf r}) \, \exp( - i E_a t) ~,
\\ 
H_a \, \varphi_a({\bf r}) & = & E_a \, \varphi_a({\bf r}) ~ ,
\end{eqnarray}
using the retarded Green function
$G( {\bf r}, t ; \, {\bf r}^\prime, t^\prime)$ which obeys  
equation
\begin{eqnarray}
\left[ i \frac{ \partial}{\partial t} - H_a - V_F(t) \right]
G( {\bf r}, t ; \, {\bf r}^\prime, t^\prime) & = &
\delta( t - t^\prime ) \, 
\delta({\bf r} - {\bf r}^\prime) 
\hspace{10mm} (t > t^\prime) ~, 
\nonumber \\
G( {\bf r}, t; \, {\bf r}^\prime, t^\prime) & = & 0
\hspace{10mm} (t < t^\prime) ~,
\end{eqnarray}
where $H_a = \frac{1}{2}{\bf p}^2 + V_a({\bf r})$ is the Hamiltonian
of an atomic system in the single active electron approximation, 
$V_a({\bf r})$ is the interaction with the core, $V_F({\bf r}, t)$ 
is the interaction with the laser field that generally includes 
the field-switching effects [below we presume that $V_F(T)$ is real].

We consider harmonic generation when the atomic system ends up
in the initial state. Other final states are also feasible, 
but till now this possibility have not been explored 
neither in the experiment,
nor in the theory. In this case the final state tends to 
$\Phi(t)_a$ for $t \rightarrow \infty$
being presented similarly to (\ref{wfg}) as
\begin{eqnarray} \label{wff}
\Psi_f( {\bf r}, t )^* & = & 
\Phi_a^*({\bf r}, t)^* + \int^{\infty}_t d t^\prime \: 
\int d^3 {\bf r}^\prime \, 
\Phi_a({\bf r}^\prime, t^\prime)^* \,
V_F({\bf r}^\prime, t^\prime) \, 
G( {\bf r}^\prime, t^\prime ; \, {\bf r}, t) ~,
\end{eqnarray} 
%The Green function (or the time-propagator) satisfies the useful equation
%\begin{eqnarray} \label{Dyson}
%G( {\bf r}, t ; \, {\bf r}^\prime, t^\prime) & = & 
%G_0( {\bf r}, t ; \, {\bf r}^\prime, t^\prime)
%+ \int^t_{t^\prime} \, 
%d t^{\prime\prime} \, \int d {\bf r}^{\prime \prime} \, 
%G_0( {\bf r}, t ; \, {\bf r}^{\prime \prime}, t^{\prime \prime}) \,
%V({\bf r}^{\prime \prime}, t^{\prime \prime}) \, 
%G( {\bf r}^{\prime \prime}, t^{\prime \prime}; \, {\bf r}^\prime, t^\prime) 
%= \nonumber \\ & = &
%G_0( {\bf r}, t ; \, {\bf r}^\prime, t^\prime)
%+ \int^t_{t^\prime} \, 
%d t^{\prime\prime} \, \int d {\bf r}^{\prime \prime} \, 
%G( {\bf r}, t ; \, {\bf r}^{\prime \prime}, t^{\prime \prime}) \,
%V({\bf r}^{\prime \prime}, t^{\prime \prime}) \, 
%G_0( {\bf r}^{\prime \prime}, t^{\prime \prime}; \, 
%{\bf r}^\prime, t^\prime) ~, 
%\end{eqnarray}
%with $G_0( {\bf r}, t ; \, {\bf r}^\prime, t^\prime)$ being the Green 
%function of the system without laser field. Note that application of 
%time-propagator $G_0$ to the bound-state function $\Phi_a$ reduces
%to plain multiplication by the phase factor according to Eq.(\ref{Psia}). 
%Employing (\ref{Dyson}) one can transform
Employing (\ref{wfg}) and (\ref{wff}) we transform $d(t)$ to 
%\begin{eqnarray} \label{dor}
%d(t) = \lim_{t^\prime , \: \: t^{\prime \prime} \rightarrow \, - \, \infty} 
%\langle \Phi_a(t^\prime) \mid G^*(t, t^\prime) \, \hat{d}_\epsilon \, 
%G(t, t^{\prime \prime}) \mid \Phi_a(t^{\prime \prime})\rangle 
%\end{eqnarray}
%to the form (cf. \cite{Buni})
\begin{eqnarray} \label{dex}
d(t) = \langle \Phi_a(t) \mid \hat{d}_{\mbox {\boldmath $\epsilon$}} 
\mid \Phi_a(t) \rangle 
+ \int^t_{- \infty} d t^\prime \,
\langle \Phi_a(t) \mid \hat{d}_{\mbox {\boldmath $\epsilon$}} \, 
G(t, t^\prime) \, V_F(t^\prime) \mid \Phi_a(t^\prime) \rangle
+ \nonumber \\ +
\int_t^{\infty} d t^\prime \,
\langle \Phi_a(t^\prime) \mid V_F(t^\prime) \, G(t^\prime, t) \,
\hat{d}_{\mbox {\boldmath $\epsilon$}} \mid \Phi_a(t) \rangle
+ \nonumber \\ +
\int_t^{\infty} dt^{\prime \prime} \,
\int^t_{- \infty} d t^\prime \, 
\langle \Phi_a(t^{\prime \prime}) \mid V_F(t^{\prime \prime}) \, 
G(t^{\prime \prime}, t) \,
\hat{d}_{\mbox {\boldmath $\epsilon$}} \, G(t, t^{\prime}) \, 
V_F(t^{\prime}) \mid \Phi_a(t^{\prime}) \rangle ~,
\end{eqnarray}
where $\langle \ldots \mid \ldots \mid \ldots \rangle$ notation is 
employed to represent integration over the space variables. 
%!
%! below
This formula could be compared with Eq.~(2.15) in Ref.~\cite{Buni}
obtained under presumption $\Psi_f \equiv \Psi_i$.
Although the general structure is similar, the important
difference lies in the range of temporal integration.
%!
In formula (\ref{dex}) the first term in the right hand side
turns zero for the inversion-invariant (for instance, spherically 
symmetric) potentials. 
Physically the second term describes the process when the 
high harmonic photon is emitted {\it after}\/ the interaction 
with the laser field, $t \ge t^\prime $,
whereas the third term describes the 'time-reversed' event
in which the radiation {\it precedes}\/ the absorption of 
laser quanta. We denote the second and the third terms
respectively as $d^+(t)$ and $d^-(t)$,
\begin{mathletters}
\begin{eqnarray} \label{dp}
d^+(t) & = & 
\int^t_{- \infty} d t^\prime \,
\langle \Phi_a(t) \mid \hat{d}_{\mbox {\boldmath $\epsilon$}} \, 
G(t, t^\prime) \, V_F(t^\prime) \mid \Phi_a(t^\prime) \rangle ~,
\\
d^-(t) & = & 
\int_t^{\infty} d t^\prime \,
\langle \Phi_a(t^\prime) \mid V_F(t^\prime) \, G(t^\prime, t) \, 
\hat{d}_{\mbox {\boldmath $\epsilon$}} \,
\mid \Phi_a(t) \rangle ~.
\end{eqnarray}
\end{mathletters}
The last term in (\ref{dex}) includes the effect usually 
referred to as {\it continuum-continuum transitions}
\cite{Hu,Lew,Buni}. In the present 
outlook it corresponds to the ``mixed'' picture when
a part of laser photons is absorbed prior to emission
of the harmonic followed by absorption of missing
low-energy quanta. 

In this paper the continuum-continuum transitions are omitted
as a rather standard approach, apparently assumed originally 
in Refs.~\cite{Hu,Lew}, albeit never scrutinized. 
There exists a simple physical reason why this term should 
not be important. The absorption of large number of low-frequency 
quanta happens when the active electron is well separated from 
an atom, see discussion of the role of large distances
in \cite{Multa,Multb}. In contrast, one should 
anticipate that 
%!absorption 
emission
of the high-energy quantum 
occurs when the electron is localized close to the atom. 
A transition of the electron from the outer 
region to the vicinity of the atom inevitably produces 
a suppression factor that describes the  electron propagation. 
Later on it will be considered in detail, see the factor $1/R$ in 
(\ref{Ma}). For a {\it natural}\/ sequence of events, when
electron first absorbs $N$ laser quanta and then emits the
high-harmonic this suppression factor appears only once. 
For continuum-continuum transitions the electron has to go 
from the outer region into the vicinity of the atom thrice, 
that induces appearance of three suppression factors, thus 
substantially reducing the amplitude. Technically the explicit 
calculation of the continuum-continuum contribution
means substantially higher level of difficulty,
since it contains {\it two}\/ Green functions;
it is not pursued in the present study. An alternative,
albeit indirect justification for omission of continuum-continuum
contribution could be seen in the fact that our 
calculations of HG rates within this approximation 
are in a good {\it quantitative}\/ agreement 
(Sec.\ref{Sec5}) with the results obtained by 
Becker {\it et al}\/ who apparently do not rely on it. 
One more argument in favor of this approximation
stems from the fact that it provides a very good agreement
between the results in length and velocity gauges,
as detailed in Sec. \ref{Sec5}.

%!   nizhe perestavleny mestami formuly
Further we presume that the monochromatic laser field
\begin{eqnarray} \label{VF}
V_F({\bf r}, t) = {\bf r} \cdot {\bf F} \, \cos \omega t
\end{eqnarray}
is switched on adiabatically at some remote time (${\bf F}$ 
is the amplitude of the electric field strength in the laser wave).
In this case one can easily check that $d^-(t) = d^+(-t)$.
After dropping in the right hand side of (\ref{dex}) \
the last term, that describes the continuum-continuum transitions, 
we arrive to the formula
\begin{eqnarray} \label{dpm}
d(t) = d^+(t) + d^+(-t) ~.
\end{eqnarray}
One should remember that in our calculations according to
(\ref{wfg}) and (\ref{wff}) we employed {\it different wave
functions}\/ $\Psi_i$ and $\Psi_f$.
Unfortunately, there is a strong trend in the literature
\cite{Eb,Hu,Lew,B,ellip,IvanovCorkum,IBB,RB,FedPet}
(to cite only part of references)
to presume that $\Psi_i \equiv \Psi_f$
that leads to replacement of the relation (\ref{dpm}) by 
\begin{eqnarray} \label{dwrong}
d(t) = d^+(t) + \left( d^+(t) \right)^* ~, 
\end{eqnarray}
and, consequently, to the real-valued dipole momentum $d(t)$.

Contrary to this almost universal delusion the expression 
(\ref{dwrong}) is incorrect 
%!   below
for description of harmonic emission by a single atom
%!   above
and the dipole momentum 
$d(t)$ {\it is not}\/ real-valued, both in the exact formulation 
(\ref{dex}) and within the approximation neglecting the
continuum-continuum transitions (\ref{dpm}).
The recent paper on the unified theory of harmonic 
%!!
%!below
generation by Becker {\it et al}\/ \cite{Buni} describes two
different approaches.
The calculations of the dipole-moment expectation
presume $\Psi_f \equiv \Psi_i$ and lead to the real-valued $d(t)$
[see Eqs.~(2.15) and (2.16) in \cite{Buni}], whereas
the $S$-matrix approach accounts for the distinction between
$\Psi_f$ and $\Psi_i$ and provides
a similar formula (2.30) but with a different range of
temporal integration [cf. our discussion below Eq.~(\ref{dex})].
The mentioned formula (2.30) gives complex-valued results.
We agree with the authors comment that the $S$-matrix approach
is to be employed when the harmonic emission of a single atom
is considered; namely this process is the subject of our present study.
However, Ref.~\cite{Buni} as
well as Refs.~\cite{Bothers,Brapid} carry
out all calculations for the dipole moment expectation
presuming that just this quantity is required as a source
term for the integration of the Maxwell equations when
the emission by the medium is considered. We believe that the
theory of collective emission should be ultimately based on
the proper description of a single atom process, but further discussion
of this issue is beyond the scope of the present paper.
%!!

For the high harmonic generation the 'time-reversed' process
described by $d^-(t)=d^+(-t)$, i.e. emission of high harmonics 
followed by absorption of a large number of the laser quanta, 
is strongly suppressed as compared with the {\it ``natural''}\/ 
sequence of events represented by $d^+(t)$, 
i.e. when at first a large number of laser quanta is gained and 
subsequently one high-frequency photon is emitted.
Therefore the further approximation of Eq.(\ref{dpm})
\begin{eqnarray} \label{dpp}
d(t) = d^+(t)
\end{eqnarray}
should work well. Below (Sec.\ref{Sec5}) we
demonstrate by explicit calculation of the Fourier components
that indeed 
%!    below
%! $\left| d^+(-t) \right| \ll \left| d^+(t) \right|$.
%!
the term $d^-(t)$ gives negligible contribution.

For monochromatic laser filed (\ref{VF})
$d(t)$ is a periodic function of time with the period
$T = 2 \pi / \omega$ where $\omega$ is the laser frequency.
Introducing the Fourier transform
\begin{eqnarray} \label{dN}
d_N = \frac{1}{T} \, \int_0^T dt \exp (i N \omega t) \, d(t) 
\end{eqnarray}
we see that 
\begin{eqnarray}
D(\Omega) = 2 \pi \sum_N \delta ( \Omega - N \omega ) \, d_N ~.
\end{eqnarray}
This {\it stationary}\/ picture does not account 
for the depletion of the initial state by the laser-induced 
transitions. To describe this effect one could employ for 
$\Psi_i$ and $\Psi_f$ the quasi-energy states with 
%!!
complex-valued quasienergies; however we do not resort 
to this complication below
%! below
(another method to account for the depletion effects was suggested
by Lewenstein {\it et al}\/ \cite{Lew}).
%! above

Thus the problem under consideration is reduced to 
calculation of the finite Fourier transform $d_N$ (\ref{dN}). 
One readily notices that
\begin{eqnarray} \label{dpmf}
d_N^+ = d^-_{-N} ~,
\end{eqnarray}
and therefore
\begin{eqnarray} \label{dnmn}
d_N = d^+_N + d^+_{-N} ~,
\end{eqnarray}
whereas (\ref{dwrong}) leads to the distinct result
\begin{eqnarray} \label{dnmnwrong}
d_N = d^+_N + \left( d^+_{-N} \right)^* ~.
\end{eqnarray}
As discussed above, for high harmonic generation we
anticipate that $\left| d^+_{-N} \right| \ll \left| d^+_{N} \right|$. 
If the term $d^+_{-N}$ is negligible, then the formulae (\ref{dnmn}) 
and (\ref{dnmnwrong}) agree. The quantitative
assessment for the quality of this approximation could be found
in Sec.\ref{Sec5}.

\subsection{Keldysh-type approximation}

Our basic approximation is to neglect the effect of the atomic core 
potential $V_a$  in the time-propagator $G(t, t^\prime)$. 
A similar assumption underlies the Keldysh \cite{Keldysh}
model, whose recent {\it adiabatic}\/ modification 
\cite{Multa,Multb,K98} gives very reliable quantitative 
results for photodetachment. A useful extension of the Keldysh
model accounts for the Coulomb electron-core interaction 
\cite{D,Krainov}. The Green function within this approximation 
is straightforwardly represented via the standard Volkov wave 
functions  $\Phi_{\bf p}({\bf r}, t)$
[$\theta(x) =1$, $(x>0)$; $\theta(x) = 0$, $(x<0)$]:
\begin{equation}\label{grfu}
G({\bf r}, t; \, {\bf r}^\prime, t^\prime)=
-i \theta( t -t^\prime)
\int  \frac{d^3 {\bf q}}{(2\pi)^3} \,  
\Phi_{\bf q}({\bf r}, t) \, 
\Phi_{\bf q}^*({\bf r}^\prime, t^\prime)  ~.
\end{equation}
Explicit expression for the Volkov functions is conveniently cast as 
\begin{eqnarray}\label{Volkov}
\Phi_{\bf p}({\bf r},t) & = &
\chi_{\bf p}({\bf r},t) \, 
\exp \left( -i \bar E_{\bf p}t \right) ~,
\\ \label{chi}
\chi_{\bf p}({\bf r},t) & = &
\exp{\left \{ i \left [ ( {\bf p}+{\bf k}_t){\bf r} -
\int_{0}^{t} \left ( E_{{\bf p}}(\tau)-
\bar E_{\bf p} \right ) 
d \tau + \frac{ {\bf p F} }{\omega^2} \right ] \right \} } ~,
\end{eqnarray}
where the factor $\chi_{\bf p}({\bf r},t)$ is time-periodic with
the period $T = 2 \pi / \omega$ , 
\begin{eqnarray}
{\bf k}_t & = & \frac{{\bf F}}{\omega} \sin \omega t ~,
\\
E_{\bf p}(t) & = & 
\frac{1}{2} \left( {\bf p} + {\bf k}_t \right)^2 ~,
\\
\bar{E}_{\bf p} & = & 
\frac{1}{T} \, \int_0^T E_{\bf p}(\tau) \, d \tau =
\frac{1}{2} p^2 + \frac{F^2}{4 \omega^2} ~.
\end{eqnarray}

Due to the property (\ref{dpmf}) it is sufficient to
restrict subsequent analysis to the $d^+_N$ component.
Using (\ref{dp}), (\ref{dN}) and (\ref{grfu}) one can rewrite 
Eq.(\ref{dN}) as 
\begin{eqnarray} \label{MGr} 
d_N & = & - \, \frac{i}{T}\int \limits_{0}^{T}dt 
\int \limits^{t}_{-\infty} dt^\prime
\int \frac{d^3 {\bf q}}{(2\pi)^3} \:
\langle \Phi_a(t) \mid e^{i \Omega t} \, 
\hat{d}_{\mbox {\boldmath $\epsilon$}} \mid \Phi_{\bf q}(t)
\rangle 
\langle \Phi_{\bf q}(t^\prime)|V_F(t^\prime)|\Phi_a(t^\prime) \rangle =
\nonumber \\
& = & - \, 
\frac{i}{T}\int \limits_{0}^{T}dt 
\int \limits^{t}_{-\infty} dt^\prime
\int \frac{d^3 {\bf q}}{(2\pi)^3} \:
\langle \varphi_a \mid \hat{d}_{\mbox {\boldmath $\epsilon$}} 
\mid \chi_{\bf q}(t) \rangle 
\langle \chi_{\bf q}(t^\prime)|V_F(t^\prime)|\varphi_a \rangle 
\times 
\nonumber \\ 
& & \times \exp \left[ i \left( E_a - \bar{E}_{\bf q} + \Omega \right) t
+ i \left(\bar{E}_{\bf q} - E_a \right) t^\prime \right] 
\end{eqnarray}
with $\Omega = N \omega$.
In the latter representation the phase factors with the phases linear
in $t$ and $t^\prime$ are explicitly singled out; the remaining
factor in the integrand is $T$-periodic both in $t$ and $t^\prime$.

\section{GREEN FUNCTION REPRESENTATION} \label{Sec3}

\subsection{Factorization technique} \label{ft}

In the present section we use the theoretical technique 
elaborated by Kuchiev \cite{K95,K96}. It allows us 
to transform the right 
hand side of (\ref{MGr}) to the form convenient for
the analysis. This transformation amounts to some special 
representation of the time-dependent Green function, 
since (\ref{MGr}) can be considered as its generalized 
matrix element. 

Generally speaking the correct description of the high 
Fourier components $d_N$ represents a formidable theoretical task. 
Its numerical implementation via solving the non-stationary
Schr\"{o}dinger equation requires both a supercomputer and exceptional 
effort. In representation (\ref{MGr}) the difficulty lies
in the strong variation of the integrand as a function of the 
time variables $t$, $t^\prime$. The application of the asymptotic
technique is hindered by the fact that in expression (\ref{MGr}) 
the integration variables are not independent: namely, the
limit of $t^\prime$-integration depends on $t$.
The crucial simplification gained by using the 
{\it factorization technique}\/ \cite{K95} 
allows us to disentangle the integration variables 
at a price of introducing an extra summation. 
Very importantly, this summation is physically meaningful as
it corresponds to the contributions of different ATI channels. 
The integration over the intermediate momenta ${\bf q}$ 
[coming from (\ref{grfu})] is carried out in closed form. 
It should be emphasized that we {\it do not}\/ use the
so called pole approximation \cite{Zar3} applied recently by Faisal 
and Becker in their model of non-sequential double ionization 
by laser field \cite{Faisal1,Faisal2}.

In order to implement this program we transform the integral 
over $t^\prime$ using the identity
\begin{equation}\label{idenfe}
\int^{t}_{-\infty}d t^\prime \exp ( i Et^\prime ) \,  
f(t^\prime) = -
i \sum_{m=-\infty}^{\infty} \frac{1}{T}\int_{0}^{T} d t^\prime
f(t^\prime) \exp \left\{ i \left[ (E - m \omega) t +
m \omega t^\prime \right] \right\}
\frac{1}{E- m \omega - i 0}~,
\end{equation}
that is valid for any periodic function $f(t)=f(t+T)$. 
The identity can be easily derived with the help of 
the Fourier expansion $f(t)=\sum_{m} f_m 
\exp ( -im\omega t )$.
Employing it we rewrite Eq.(\ref{MGr}) in the form 
of the series
\begin{eqnarray} \label{Mfin}
d^+_N = \sum_m d^+_{N \, m} ~,
\end{eqnarray}
where
\begin{eqnarray}\label{Aden}
d^+_{N \, m} = & - \, 
& \frac{1}{T^2}\int_{0}^{T}dt \int_{0}^{T} dt^\prime
\int \frac{d^3 {\bf q}}{(2\pi)^3} \,
\frac{\langle \varphi_a \mid \hat{d}_{\mbox {\boldmath $\epsilon$}} 
\mid \chi_{\bf q}(t)\rangle 
\langle \chi_{\bf q}(t^\prime)|V_F(t^\prime)|\varphi_a\rangle }
{\bar{E}_{\bf q} - E_a - m \omega - i0} \times
\\ \nonumber
& \times & \exp \left[
%E_a - \bar{E}_{\bf q} +\bar E_c-E_b - E_a - m \omega) t + 
i (\Omega - m \omega) t + i m \omega t^\prime \right] ~.
\end{eqnarray}

The next step is to carry out the integration over ${\bf q}$. 
To this end we note that the wave function 
$\chi_{{\bf q}}({\bf r},t)$ [see  Eq.(\ref{chi})],
depends on ${\bf q}$ in a very simple way. Namely, it is 
an exponent of a linear form of ${\bf q}$: 
$\chi_{{\bf q}}({\bf r},t) = \exp \left\{ i {\bf q} 
\left[ {\bf r} + ({\bf F}/\omega^2) \, \cos \omega t \right]
+ \alpha \right\}$,
where $\alpha$ is ${\bf q}$-independent phase. 
The other source of ${\bf q}$ dependence in the integrand is its 
denominator that contains term $\frac{1}{2}q^2$. The calculation of 
three-dimensional integrals of this type is standard.
%!
%! formuly prestavleny mestami
%!
The integration over $|{\bf q}|$ might be carried out as a contour
integration in the complex plane in order to specify the following 
validity conditions:
\begin{eqnarray} \label{ReR}
{\rm Re} R > 0 ~,
\\ \label{ImK}
{\rm Im} K_m > 0 ~,
\end{eqnarray}
%$Im K_m > 0$ 
where 
\begin{equation}\label{Kmnm}
K_{m} =\sqrt{ 2 \left ( m \omega -
\frac{F^2 }{4 \omega^2} + E_a\right ) }
\end{equation}
is the photoelectron momentum after absorption of $m$ 
laser quanta (see also Sec.\ref{Sec3B}),
$R = \sqrt{{\bf R}^2}$ is the function of all the variables of integration 
\begin{equation}\label{R12}
{\bf R}={\bf R}({\bf r},{\bf r}^\prime;t,t^\prime)=
\int_{t}^{t^\prime} {\bf k}_{\tau} d\tau - {\bf r}^\prime +{\bf r} 
=\frac{{\bf F}}{\omega^2} \left( \cos \omega t -
\cos \omega t^{'} \right) + {\bf r} - {\bf r}^\prime ~.
\end{equation}
Note that below we shift the integration to the complex-valued 
time $t^\prime$, and hence complex-valued $R$.
After integration over ${\bf q}$ the expression for $d^+_{N \, m}$ reads
\begin{eqnarray}\label{Mden}
d^+_{N \, m} = - \, \frac{1}{T^2}\int_{0}^{T}dt \int_{0}^{T} dt^\prime
\int d^3 {\bf r} \, \int d^3 {\bf r}^\prime \, 
\varphi_a({\bf r}) \, 
({\bf \epsilon} \cdot {\bf r} ) \, 
({\bf r}^\prime \cdot {\bf F} \cos \omega t^\prime) \,  
\varphi_a({\bf r}^\prime)
\times  
\nonumber \\ \times 
%\left ( -\frac{1 }{2 \pi R} \right )
\frac{1}{2 \pi R} \, 
\exp \left\{ i \left [K_m R + 
(\Omega  - m \omega) t + m \omega t^\prime +
{\bf k}_{t}{\bf r}-
{\bf k}_{t^\prime}{\bf r}^\prime +
\int^{t}_{t^\prime} d \tau
\left ( \frac{1}{2}{\bf k}_{\tau}^2- \frac{F^2}{4\omega^2} \right )
\right ] \right\} ~. 
\end{eqnarray}

Eq.(\ref{Mden}) is convenient for evaluating the parameters
governing the process. First of all note that integration
over the variables ${\bf r}$, ${\bf r}^\prime$ is localized in
the vicinity of the atom. 
%Indeed, integration over ${\bf r}^\prime$ 
%describes the matrix elements responsible for the multiphoton
%ionization from the initial bound state, whereas the integration 
%over ${\bf r}$ describes the matrix elements of the free-bound 
%transition with emission of the quantum $\Omega$.
The characteristic atomic dimensions should be compared with
the amplitude $F/ \omega^2$ of the electron wiggling in 
the laser field (the latter motion being described by
$\int^t {\bf k}_\tau \, d \tau$).
This amplitude becomes large even for quite moderate electric fields
\begin{equation}\label{fom}
\frac{F}{\omega^2} \gg 1~.
\end{equation}
Note that this inequality may be satisfied both for 
large ($\gamma >1$) as well as small ($\gamma <1$) values 
of the Keldysh adiabaticity parameter
$\gamma = \omega \sqrt{ 2 | E_a |}/F$.
Therefore, for the fields satisfying (\ref{fom}) we can assume that 
\begin{equation}\label{abbrr}
\frac{F}{\omega^2} \gg  r,r^\prime ~.
\end{equation}
Actually the applicability of the resulting approximation is 
even broader than outlined above since in fact relation 
(\ref{fom}) has to be replaced by its more accurate version
\begin{equation}\label{fomm}
\frac{( - e F)}{\omega^2} 
\mid \cos \omega t - \cos \omega t^\prime \mid \gg 1 ~.
\end{equation}
The main contribution corresponds to the complex values of
time when $| \cos \omega t^\prime | \gg 1$, as elaborated 
in the adiabatic approximation of Sec.\ref{Secad} .

These observations allow us to simplify $R$ by neglecting 
${\bf r}$ and ${\bf r}^\prime$ in the pre-exponential factor 
of the integrand in (\ref{Mden}) and retaining the first order 
terms in the phase:
\begin{eqnarray}\label{RR0}
& &R\approx R_0 + \frac{ {\bf R}_0\cdot
({\bf r}-{\bf r}^\prime)}{  R_0 } ~,
\\ \label{R0sq}
& &R_0=R_0 (t,t^{\prime})= \sqrt{{\bf R}_0^2} ~,
\\ \label{R0tt}
& &{\bf R}_0 ={\bf R}_0(t,t^\prime) =
\frac{({\bf F})}{\omega^2}(\cos \omega t-\cos \omega t^{\prime}) ~.  
\end{eqnarray}
As a result the Green function in (\ref{Mden}) becomes simpler
in the 'wave zone':
\begin{eqnarray}\label{GR0}
& &\frac{1}{R} 
\exp \left( i K_m R \right) \approx
\frac{1}{R_0} 
\exp \left\{ i \left[ K_m R_0 +  {\bf K}_m \cdot
\left({\bf r} - {\bf r}^\prime \right) \right] \right\} ~,
\\ \label{KFF}
& &{\bf K}_m = \sigma K_m \frac{{\bf F}}{F} ~,
\quad \quad \quad \sigma = \pm 1 ~.
\end{eqnarray}
Substituting Eq.(\ref{GR0}) in (\ref{Mden}) and using
Eq.(\ref{Kmnm}) to rewrite the exponent we find
\begin{eqnarray}\label{facrr}
d^+_{N \, m} = & - & \, \sum_\sigma \,
\frac{1}{ 2 \pi T^2}\int_{0}^{T}dt \int_{0 \, ({\cal C})}^{T} dt^\prime
\int d^3 {\bf r} \, \int d^3 {\bf r}^\prime \, 
\varphi_a({\bf r}) \, 
({\bf \epsilon} {\bf r} ) \, 
({\bf r}^\prime {\bf F} \cos \omega t^\prime) \,  
\varphi_a({\bf r}^\prime) \,
\times  
\nonumber \\ & \times &
\frac{1}{R_0(t, t^\prime)}
%\frac{1}{R_0(t, t^\prime)} \, 
\exp \left\{ i \left [K_m R_0(t, t^\prime) + 
(\Omega  - m \omega) t + m \omega t^\prime +
\left( {\bf K}_m + {\bf k}_{t} \right) {\bf r}-
\right. \right. 
\nonumber \\ 
& - & \left. \left.
\left( {\bf K}_m + {\bf k}_{t^\prime} \right) {\bf r}^\prime +
\int^{t}_{t^\prime} d \tau
\left ( \frac{1}{2}{\bf k}_{\tau}^2
- \frac{{\bf F}^2}{4\omega^2} \right )
\right ] \right \} ~.
\end{eqnarray}
Here summation over $\sigma = \pm 1$ indicates that one has to
take ${\bf K}_m$ parallel or antiparallel to ${\bf F}$ depending on 
the sign of $(\cos \omega t - \cos \omega t^\prime)$ in order to
satisfy the convergence condition (\ref{ReR}) (we postpone till
the next section the more detailed discussion of this issue).  
Note that the denominator $R_0$ in (\ref{facrr}) can turn zero thus 
challenging convergence of integrals. This problem is circumvented if
one presumes that the integration over $t^\prime$ is shifted
from the real axis into the upper half-plane of complex-valued time. 
The integration contour ${\cal C}$ will be specified more exactly below.

Now we return to the time-dependent bound state wave function
$\Phi_a({\bf r}, t)$ and Volkov states $\Phi_{\bf p}({\bf r}, t)$
bearing in mind relation (\ref{Kmnm}). 
%For the preliminary discussion we make a plain simplifying presumption 
%that 
%$R_0(t, t^\prime) = (-eF/\omega^2) \left( \cos \omega t - 
%\cos \omega t^\prime \right)$, i.e.
%${\rm Re} \left( \cos \omega t - \cos \omega t^\prime \right) > 0$
%[see the condition (\ref{ReR})], postponing detailed discussion
%till the next section. 
This gives us an appealing representation: 
\begin{eqnarray} \label{Mpref}
d^+_{N \, m} = & - & \, \sum_\sigma \,
\frac{1}{2 \pi T^2}\int \limits_{0}^{T} dt 
\int \limits_{0 \, ({\cal C})}^{T}dt^\prime \,
\frac{1}{ R_0(t,t^\prime) }
\times \nonumber \\ 
& \times & 
\langle \Phi_a(t) \mid \exp ( i \Omega t ) \, 
\hat{d}_{\mbox {\boldmath $\epsilon$}} \mid 
\Phi_{{\bf K}_m} (t) \rangle \langle 
\Phi_{{\bf K}_m} (t^\prime) \mid V_F(t^\prime) \mid
\Phi_a (t^\prime) \rangle ~.
\end{eqnarray}

\subsection{Above Threshold Ionization and Laser Assisted 
Recombination} \label{Sec3B}

An interpretation of formula (\ref{Mpref}) is based on 
the fact that the integrand is a product of physically
meaningful factors. In order to recognize the first of 
them one should recall that the amplitude of $m$-photon  
detachment of electron from the initial state $\Phi_a$ within 
the Keldysh \cite{Keldysh} approximation is given by
\begin{eqnarray} \label{AKeld}
A_m({\bf p}) = \frac{1}{T}\int \limits_{0}^{T} dt \,
\langle \Phi_{{\bf p}} (t) \mid V_F(t) \mid \Phi_a (t) \rangle ~ .
\end{eqnarray}
Generally, the number $m$ of photons absorbed is larger than
a minimum necessary for the electron detachment, therefore
the process corresponds to the Above Threshold Ionization
(ATI). In the right hand side of (\ref{AKeld}) the index 
$m$ is implicit. It enters via the absolute value of the 
final electron momentum ${\bf p}$ which is subject to 
the energy conservation constraint 
\begin{eqnarray} \label{ec}
\frac{1}{2} p^2 = m \omega - \frac{F^2}{4 \omega^2} + E_a ~.
\end{eqnarray}
Note that $-E_a \equiv \frac{1}{2} \kappa^2 > 0 $ is the electron 
binding energy in the initial state; $F^2/(4 \omega^2)$ 
is the electron quiver energy in the laser field.
Comparing (\ref{ec}) with (\ref{Kmnm}) we confirm that $K_m$ 
is exactly the physical translational electron momentum 
in the $m$-th ATI channel. 

The second relevant process is that of laser assisted 
recombination (LAR) when the continuum electron with 
the momentum ${\bf p}$ collides with an atom in the laser 
field. The electron absorbs some extra laser quanta 
and goes to the bound state $\Phi_a$ emitting 
single photon of frequency $\Omega$. 
In the Keldysh-type approximation this LAR process has 
the amplitude 
%%% Question sign minus  %%%
\begin{eqnarray} \label{rec}
C_{N \, m}({\bf p}) = -\frac{1}{2 \pi T}\int \limits_{0}^{T} dt \,
\langle \Phi_a (t) \mid \exp ( i \Omega t ) \, 
\hat{d}_{\mbox {\boldmath $\epsilon$}} \mid
\Phi_{\bf p} (t) \rangle ~. 
\end{eqnarray}
It could be called also {\it laser-induced 
recombination}. The allowed values of the high-energy 
photon frequency $\Omega$ are
\begin{eqnarray}
\Omega = \frac{1}{2} p^2 + \frac{F^2}{4 \omega^2}
+ m \omega
\end{eqnarray}
with integer $m$.

One readily notices that the integrand in (\ref{Mpref}) 
bears a striking resemblance to the product of the integrands 
in (\ref{AKeld}) and (\ref{rec}). 
This allows us to interpret the amplitude (\ref{Mpref}) 
as describing the two-step transition:
at first the electron goes to the laser-field dressed continuum 
(Volkov) state $\Phi_{{\bf K}_m}$ after absorption of $m$ photons; 
at the second step it returns to the initial state emitting 
single photon with the frequency $\Omega = N \omega$. 
Note that the intermediate momentum ${\bf p} = {\bf K}_m$
is restricted not only in magnitude (\ref{Kmnm}) but also 
in direction: according to Eq.(\ref{KFF}) the vector ${\bf K}_m$ 
is parallel or anti-parallel to the electric vector ${\bf F}$ 
in the laser wave. The physical implications of this circumstance 
are discussed in the next section.

There is an extra factor $1/R(t, t^\prime)$ in the integrand 
of (\ref{Mpref}) which prevents complete separation of 
integrations in $t$ and $t^\prime$. In fact its presence
is physically well understandable. Indeed, the definition 
of $R(t, t^\prime)$ (\ref{R12}) shows that
classically it is the distance between 
the electron positions at the moments $t$ and $t^\prime$ 
with account for the electron wiggling in the laser field.
$1/R$ could be named an {\it expansion factor}\/
since in the absence of the laser field ($F \rightarrow 0$)
it describes conventional decrease of the amplitude in a
spherical wave as it expands in 3D space
($\sim 1/| {\bf r} - {\bf r}^\prime |$). When the laser 
field is operative, the form of the expansion factor is 
drastically modified according to the approximation (\ref{R0tt}).
                                                                                Hence the interpretation of the expression (\ref{Mpref}) 
is that the electron first is transferred to the $m$-th ATI channel, 
then propagates in space under the influence of the 
laser wave and finally recombines to the initial state emitting
the photon with the frequency $\Omega$.
The contribution of each path is labeled by the number of
virtually absorbed photons $m$. These contributions add up 
coherently as shown by Eq.(\ref{Mfin}).
For the actual ATI process the momentum $K_m$ should be real. 
The summation in (\ref{Mfin}) includes also 
the virtual processes with the imaginary values of $K_m$, 
but their contribution is anticipated to be small.

It is worthwhile now to outline equivalent but more 
convenient representation of ATI and LAR amplitudes.
Bearing in mind the character of ${\bf r}$-dependence of Volkov states
(\ref{Volkov})-(\ref{chi}) it is easy to see that the space
integration in the formulae (\ref{AKeld}) or (\ref{rec}) 
essentially reduces to the Fourier transformation.
For instance, as shown in much detail by Gribakin and Kuchiev \cite{Multa},
the ATI amplitude (\ref{AKeld}) is exactly presented as
\begin{eqnarray} \label{AF}
A_m({\bf p}) = - \frac{1}{2T}\int \limits_{0}^{T} dt 
\left[ \left( {\bf p} + {\bf k}_t \right)^2 + \kappa^2 \right]
\tilde{\phi}_a \left( {\bf p} + {\bf k}_t \right)
\exp \left[ i S(t) \right] ~ ,
\end{eqnarray}
where $S(t)$ is the classical action
\begin{eqnarray}
S(t) = \frac{1}{2} \int^t d \tau \left({\bf p} + {\bf k}_{\tau} \right)^2
-E_a t ~,
\end {eqnarray}
$\tilde{\phi}_a({\bf q})$ is the Fourier transform of $\phi_a({\bf r})$:  
\begin{eqnarray} \label{wF}
\tilde{\phi}_a({\bf q}) = \int d^3 {\bf r} \, 
\exp( - i {\bf q} {\bf r} ) \, \phi_a({\bf r}) ~ .
\end {eqnarray}
%and $\kappa = \sqrt{ 2 |E_a|}$.
%
%Along the same line the harmonic generation amplitude (\ref{Mpref})
%is cast as 
%\begin{eqnarray} \label{MF}
%{\cal M}_{N \, m}(\Omega) = & &
%\frac{1}{8 \pi T^2}\int \limits_{0}^{T} dt 
%\int \limits_{0 \, ({\cal C})}^{T}dt^\prime \,
%\frac{1}{ R_0(t,t^\prime) } \,
%\left[ \left( {\bf K}_m + {\bf k}_t \right)^2 + \kappa^2 \right]
%\left[ \left( {\bf K}_m + {\bf k}_{t^\prime} \right)^2 + \kappa^2 \right]
%\times \nonumber \\ 
%& & \times  
%\tilde{\Phi}_a \left( {\bf K}_m + {\bf k}_t \right) \, 
%\tilde{\Phi}_a \left( {\bf K}_m + {\bf k}_{t^\prime} \right) \, 
%\exp \left\{ i \left[S(t^\prime) - S(t) + \Omega t \right] \right\} ~.
%\end{eqnarray}
Similarly, for the LAR amplitude one obtains 
\begin{eqnarray} \label{CF} 
C_{N \, m} = - \, \frac{1}{2 \pi T} \, \int \limits_{0}^{T} dt \:
\exp \left\{ i \left[\Omega t - S(t) \right] \right\} \:
\tilde{\phi}_a^{({\mbox {\boldmath $\epsilon$}})} 
\left( - {\bf K}_m - {\bf k}_t \right) 
\end{eqnarray}
with
\begin{eqnarray} 
\tilde{\phi}_a^{({\mbox {\boldmath $\epsilon$}})}({\bf q}) = i 
\left( {\mbox {\boldmath $\epsilon$}} 
\cdot \nabla_{\bf q} \right) \tilde{\phi}_a({\bf q}) ~. 
\end{eqnarray}

Before concluding this section it is worthwhile to make an important
observation. It is well known that the Keldysh approximation in 
the theory of multiphoton ionization is not gauge invariant.
%!
%!   below
The detailed discussion of this issue can be found in Ref.~\cite{Sh}.
%!The same applies to the present approach to the harmonic generation
%!problem.
%!   above
Basing on physical grounds we use the dipole-length form
(\ref{VF}) for the interaction of the electron with the laser field.
As thoroughly discussed earlier \cite{Multa,Multb}, 
this gauge stresses large separations of the active electron 
from the core where one-electron approximation is better justified 
and one can employ  the asymptotic form for the initial-state wave 
function:
\begin{eqnarray} \label{wf}
\phi_a({\bf r}) \approx A_a r^{\nu-1} \, \exp(- \kappa r) \,
Y_{lm}(\hat{{\bf r}})
\quad \quad \quad
( r \gg 1/\kappa),
\end{eqnarray}
where $\kappa = \sqrt{ 2 |E_a|}$, $\nu = Z/\kappa$, $Z$
is the charge of the atomic residual core ($\nu=Z=0$ for a
negative ion), $l$ is the active electron orbital momentum in
the initial state and $\hat{{\bf r}} = {\bf r}/r$ is the unit vector.
The coefficients $A_a$ are tabulated for many negative ions \cite{RS}.
The Fourier transform $\tilde{\Phi}_a({\bf q})$ (\ref{wF})
is singular at $q^2 = \kappa^2$ with the asymptotic behavior for 
$q \rightarrow \pm i \kappa$ defined by the long-range asymptote
(\ref{wf}) in the coordinate space
\begin{eqnarray} \label{wfmo}
\tilde{\phi}_a({\bf q}) = 4 \pi A_a (\pm 1)^l \, Y_{lm}(\hat{{\bf q}}) \,
\frac{(2 \kappa)^\nu \, \Gamma(\nu + 1)}{(q^2 + \kappa^2)^{\nu +1}} ~, 
\end{eqnarray}
with $(\pm 1)^l$ corresponding to $q \rightarrow \pm i \kappa$. 

At the present stage of our development the interpretation of
(\ref{Mpref}) outlined above should be considered only as 
qualitative, since the two time integrations in Eq.(\ref{Mpref}) 
are not completely separated due to the factor $1/R_0(t, t^\prime)$. 
The complete factorization is possible under additional approximation
which actually is not restrictive possessing a broad applicability range. 

\section{ADIABATIC ANALYSIS} \label{Secad}

\subsection{Adiabatic approach to Above Threshold Ionization}
\label{AdATI}

We start with reiterating the basics of 
the adiabatic approximation in multiphoton detachment theory 
\cite{Multa,Multb} that allows one to carry out 
analytically integration in expression
(\ref{AKeld}) or (\ref{AF}) for the amplitude. It is presumed 
that the laser frequency $\omega$ is small, i.e. that the number 
of absorbed photons $m$ is large (the practical applicability 
of this approach proves to be very broad, since it gives 
reasonable results even for $m = 2$, see, for instance, 
Ref.\cite{K98}). 
Then the integrand in (\ref{AKeld}) or (\ref{AF}) 
contains large phase factor
$\exp \left[ i S(t) \right]$
and the integral may be evaluated using the saddle point method 
\cite{Multa,Multb,K98,K99}. 
The positions of the saddle points in 
the complex $t$ plane is defined by equation 
\begin{eqnarray} \label{seqa}
S^\prime(t_{m \mu}) = 0 ~,
\end{eqnarray}
or, more explicitly,
\begin{eqnarray} \label{seqb}
\left( {\bf p} + {\bf k}_{t_{m \mu}} \right)^2 + \kappa^2 \equiv 
\left( {\bf p} + \frac{{\bf F}}{\omega} 
\sin \omega t_{m \mu} \right)^2 + \kappa^2 = 0 ~.
\end{eqnarray}
Eqs.(\ref{seqa}) or (\ref{seqb}) are to be considered 
together with the energy conservation constraint Eq.(\ref{ec}).
Note that according to formula (\ref{wfmo}) the position 
of the saddle point coincides with the singularity 
of the bound-state wave function in the momentum space thus 
stressing the importance to describe correctly the 
long-range behavior of coordinate-space wave function. 
The latter could be ensured much more easily than 
the proper description of the wave function inside 
the core region. Therefore 
the use of the adiabatic approximation and characterization 
of the bound wave function solely by its asymptotic 
behavior (\ref{wf}) constitutes a self-consistent 
approach. 
%Parenthetically it could be noted that 
%application of the adiabatic approximation to the LAR
%amplitude (that is not pursued below) is not as well 
%justified just because in this case the related 
%saddle point does not coincide with the singularity 
%of the bound state wave function in the momentum space.

The result of calculations of the amplitude (\ref{AKeld}) in the
stationary phase approximation could be written as a modification 
of formula (25) in 
%the paper by Gribakin and Kuchiev
Ref.\cite{Multa}:
\begin{eqnarray} \label{A}
A_m^{(sp)}({\bf p}) & = & \sum_{\mu} \, A_{m \, \mu}^{(sp)}({\bf p}) ~, 
\\ \label{Amu}
A_{m \, \mu}^{(sp)}({\bf p}) 
& = & - \frac{(2 \pi)^2}{T} \, A_a \, \Gamma(1+\nu/2)
\, 2^{\nu/2} \, \kappa^\nu \, Y_{lm}(\hat{{\bf p}}_{ m\mu}) \,
\frac{\exp \left[ i S(t_{m \mu}) \right]}
{\sqrt{2 \pi \left[- i S^{\prime \prime}(t_{m \mu})^{\nu+1} \right]}} ~.
\end{eqnarray}
In the plane of the complex-valued time the saddle points 
$t_{m \mu}$ lie symmetrically with respect to the real axis. 
There are four saddle points in the interval 
$ 0 \leq {\rm Re} \, t_{m \mu} \leq T$, two of them lying 
in the upper half plane (${\rm Im} \, t_{m \mu} > 0$).
The integration contour in the plane of complex time is shifted 
upwards. Therefore only two saddle points with 
${\rm Im} \, t_{m \mu} > 0$ are operative 
being included into the summation in (\ref{A})
($\mu = \pm 1$);  $\hat{{\bf p}}_{m \mu}$ is a unit vector in 
the direction of ${\bf p} + {\bf k}_{t_{m \mu}}$.

\subsection{Adiabatic approach to Harmonic Generation}

Although in the harmonic generation matrix element (\ref{Mpref})
the time integration variables $t$ and $t^\prime$ are not fully 
separated, as indicated above, the large phase factor 
$\exp \left[ i S(t^\prime) \right]$ in $t^\prime$ integration 
is the same as in the multiphoton detachment case (\ref{AKeld}). 
Therefore we can apply the saddle point approximation to carry
out integration over $t^\prime$. The integration contour is
again shifted upwards in the complex $t^\prime$ plane; this
is the contour ${\cal C}$ in the formulae (\ref{facrr}) and 
(\ref{Mpref}). The saddle points are defined from the same 
Eq.(\ref{seqb}). By solving it we find for the 
$m$-dependent  saddle points $t_{m \, \mu}^\prime$:
\begin{eqnarray} 
\sin \omega t_{m \mu}^\prime & = & \frac{\omega}{F} 
\left(- K_m + i \mu \kappa \right) ~. 
\nonumber \\
\cos \omega t_{m \mu}^\prime & = &
\sqrt{1- \left(\frac{\omega}{F} \right)^2 \, 
\left( - K_m + i \mu \kappa \right)^2 } 
\hspace{10mm} \left( {\rm Im} \cos \omega t_{m \mu}^\prime > 0 \right) ~.
\end {eqnarray}
For each value of $m$ there are two essential saddle points,
i.e. these with ${\rm Im} \, t_{m \, \mu}^\prime > 0$, 
labeled by subscript $\mu = \pm 1$.

Thus in the stationary phase approximation for the $t^\prime$ 
integration we obtain from (\ref{Mpref})
\begin{eqnarray} \label{MFsp}
d_{N \, m}^{+}(\Omega) & = & \sum_\sigma \, \sum_{\mu} \, 
A_{m \, \mu}^{(sp)}({\bf K}_m) \, B_{N \, m \mu} ~,
\\ \label{B}
B_{N \, m \mu} & = & - \,  
\frac{1}{2 \pi T}\int \limits_{0}^{T} dt \: 
\frac{1}{ R_0(t,t_{m \mu}^\prime) } \,
\langle \Phi_a(t) \mid \exp ( i \Omega t ) \, 
\hat{d}_{\mbox {\boldmath $\epsilon$}} \mid 
\Phi_{{\bf K}_m} (t) \rangle ~.
\end{eqnarray}
Similarly to Eq.(\ref{CF}) one can conveniently employ 
the Fourier transform which gives
\begin{eqnarray} \label{BF} 
B_{N \, m \mu} = - \, \frac{1}{2 \pi T} \, \int \limits_{0}^{T} dt \:
\frac{\exp \left\{ i \left[\Omega t - S(t) \right] \right\} }
{( F / \omega^2) 
\left( \cos \omega t - \cos \omega t^\prime_{m \mu} \right) } \:
\tilde{\phi}_a^{({\mbox {\boldmath $\epsilon$}})} 
\left( - {\bf K}_m - {\bf k}_t \right) ~. 
\end{eqnarray}
In particular, for a negative ion ($\nu = 0$) with the active 
electron in an $s$ state ($l=0$) we have from (\ref{wfmo}) 
($ \hat{{\bf q}} \equiv {\bf q}/q$)
\begin{eqnarray}
\tilde{\phi}_a({\bf q}) & = & \sqrt{4 \pi} A_a 
\frac{1}{(q^2 + \kappa^2)} ~, 
\\
\tilde{\phi}_a^{({\mbox {\boldmath $\epsilon$}})}({\bf q}) & = & - i  
\left( {\mbox {\boldmath $\epsilon$}} \cdot \hat{{\bf q}} \right)
\sqrt{4 \pi} A_a \frac{2 q}{(q^2 + \kappa^2)^2} ~, 
\end{eqnarray}
and (\ref{BF}) simplifies to 
\begin{eqnarray} \label{Bs}
B_{N \, m \mu} = i \,
\frac{2 A_a}{\sqrt{\pi} T} \,
\frac{\omega^2}{ F } \,
\int \limits_{0}^{T} dt \:
\frac{\exp \left\{ i \left[\Omega t - S(t) \right] \right\}}
{ \cos \omega t - \cos \omega t^\prime_{m \, \mu} } \:
\frac{ {\mbox {\boldmath $\epsilon$}} \cdot 
\left( {\bf K}_m + {\bf k}_t \right)}
{\left[ \left( {\bf K}_m + {\bf k}_t \right)^2 + \kappa^2 \right]^2} ~.
\end{eqnarray}

%Apparently the integration over $t$ variable in (\ref{Bs}) also 
%might be carried out by the saddle point method. 
%The saddle point is given here by
%\begin{eqnarray} \label{sprad}
%\sin \omega t_{m \, \mu} = \frac{\omega}{- e F} 
%\left(K_m + \mu \sqrt{\left(2 \Omega - \kappa^2 \right)} \right) ~. 
%\end {eqnarray}
%and
%\begin{eqnarray} \label{Bsp}
%B_{m \, \mu}^{(sp)} & = &
%\frac{1}{4 \pi T} \,
%\frac{1}
%{( - e F / \omega^2) 
%\left(\cos \omega t^\prime_{m \, \mu}  
%- \cos \omega t_{m \, \mu} \right) } \,
%\left[ \left( {\bf K}_m + {\bf k}_t \right)^2 + \kappa^2 \right]
%\times \nonumber \\ 
%& \times & 
%\tilde{\Phi}_a \left( \sqrt{\kappa^2 - 2 \Omega} \right) \, 
%\exp \left\{ i \left[- S(t) + \Omega t \right] \right\} ~.
%\end{eqnarray}
%Note that in this case the saddle point position defined 
%differs from the pole of $\tilde{\Phi}_a({\bf q})$. 
%Hence the integral over $t$ variable {\it is not defined}\/ by 
%the long-range asymptote of $\Phi_a({\bf r})$.
%In practice our calculations have shown that the saddle point
%approximation gives good results for calculation of the integral 
%of the form (\ref{Bs}) but modified to retain only the phase factor
%$\exp \left\{ i \left[- S(t) + \Omega t \right] \right\}$
%in the integrand. However if the preexponential factor is
%present, as it is necessary for the actual calculations,
%the results become unacceptably bad. We did not explore further
%the origin of this failure, but merely have resort to the 
%numerical evaluation of the integral (\ref{Bs}) in our calculations 
%discussed below. 

As noted above Eq.(\ref{Mpref}), the summation over $\sigma = \pm 1$
indicates that in order to satisfy the convergence condition (\ref{ReR})
one has to choose a sign in
$R_0(t, t^\prime) = \pm (F/\omega^2) \left( \cos \omega t - 
\cos \omega t^\prime \right)$
that makes $R_0(t, t^\prime) > 0$.
By considering the phase factor in (\ref{facrr}), we see 
that the upper sign corresponds to 
${\bf K}_m$ parallel to ${\bf F}$ 
whereas the lower sign leads to ${\bf K}_m$ antiparallel to 
${\bf F}$ (without changing $|{\bf K}_m|$). This looks
natural since there is no reason to prefer one of these 
directions. 

The right hand side of formula (\ref{MFsp}) contains also 
summation over the saddle points $t^\prime_{m \, \mu}$ 
(for the monochromatic laser field two such points are operative 
being labeled by the index $\mu = \pm 1$).
However, as argued by Kuchiev \cite{K95}, actually only one saddle 
point (i.e. one value of the label $\mu$) contributes 
for each choice of $\sigma = \pm 1$, 
i.e. for ${\bf K}_m$ parallel to ${\bf F}$ and 
for ${\bf K}_m$ antiparallel to ${\bf F}$
(as discussed below, this has a simple and clear physical
interpretation). 
Consequently the double summation over $\sigma$ and $\mu$
is effectively replaced by a single summation.
Moreover, in the latter sum both terms are equal.
To see this one should remember that the preexponential factor
$1/R_0(t, t^\prime)$ changes sign depending on the value
of $\sigma = \pm 1$. Hence the remaining single summation
is equivalently replaced by the factor of 2. This allows us
to finally rewrite (\ref{Mfin}) using (\ref{MFsp}) as
\begin{eqnarray} \label{dfin}
d^+_N = 2 \, \, \sum_m \, A_{m \, \mu_0}^{(sp)}({\bf K}_m) 
\, B_{N \, m \mu_0} ~, 
\end{eqnarray}
where $A_{m \, \mu_0}^{(sp)}({\bf K_m})$ and $B_{N \, m \mu_0}$ 
are given respectively by (\ref{Amu}) and (\ref{Bs}) for 
negative ion ($\nu = 0$). 
Expression (\ref{dfin}) is to be calculated for the
subscript $\mu$ corresponding to one of the saddle points 
$t^\prime_{m \: \mu = \mu_0}$, for instance, that with the smaller
value of ${\rm Re} \, t^\prime_{m \mu_0}$ 
(and ${\rm Im} \, t^\prime_{m \mu_0} > 0$).

In formula (\ref{dfin}) the factor $B_{N \, m \mu_0}$
describes jointly the 3D-wave expansion and LAR. These two effects
could be further factorized using the approximation
$|\cos \omega t^\prime_{m \, \mu_0}| \gg |\cos \omega t |$  
\cite{K95}:
%!!   formula added below 
\begin{eqnarray} \label{Bsimpl}
&& B_{N \, m \mu_0} =  \frac{1}{R_{m \mu_0}} \, 
C_{N \, m}({\bf K}_m) ~,
\\ \label{Ma}
&&d^+_N = 2 \, \, \sum_m \, A_{m \, \mu_0}^{(sp)}({\bf K}_m) 
\, \frac{1}{R_{m \mu_0}} \, 
C_{N \, m}({\bf K}_m) ~,
\end{eqnarray}
where $1/R_{m \mu_0} = \omega^2 /(F \cos t_{m \mu_0}^\prime)$ 
is the laser-modified expansion factor in its simplest form
and $C_{N m}$ (\ref{CF}) is LAR amplitude.  

The adiabatic approximation could be further applied 
to carry out integration over $t$ in $B_{N \, m \mu}$ 
(\ref{Bs}) or in $C_{N m}$ (\ref{CF}) by the saddle point
method. However below we do not pursue this objective
and evaluate these integrals numerically.

Eq.(\ref{dfin}) and its simplified version (\ref{Ma})
present the main result of this paper. These formulae 
implement the very simple picture of the HG process
as consisting of three successive steps.
First, the electron  absorbs $m$ laser photons.
The amplitude of this event is  $A_{m \, \mu} $.
In order to contribute to HG the photoelectron has to 
return to the parent atomic core where LAR is solely 
possible. The amplitude of return is described by the expansion
factor $1/R$. It appears explicitly in (\ref{Ma}), while
in (\ref{dfin}) it is incorporated in the definition of the
amplitude $B_{N \, m \mu_0}$.
The propagation of the electron describes the second
phase of the event.
At the third step the electron collides
with the core absorbing $N-m$ photons from 
the laser field and emitting the single high-frequency 
quantum $\Omega = N \omega$ as it recombines to the bound state. 
This LAR process has the amplitude $C_{Nm}$ (\ref{rec}).
The summation over $m$ in the total amplitude $d_N$ 
(\ref{dfin}) takes into account interference
of the transitions via different intermediate ATI channels.

%!
This appealing physical picture is supplemented by the very
simple way to evaluate numerically all the quantities in
(\ref{dfin}). The amplitude of photoionization
$A_{m \, \mu_0}^{(sp)}$ is calculated via plain analytical
formulae with the validity well testified before.
The LAR process did not attract much attention in the literature
and certainly deserves more study that we hope to present 
elsewhere. Here we emphasize only that since 
$C_{N \, m \mu_0}$ as well as closely related amplitude
$B_{N \, m \mu_0}$ are very similar in structure
to the amplitude $A_{m \, \mu_0}^{(sp)}$, one
can hope that similar methods of evaluation also produce 
reliable results.

The nontrivial point 
in the presented picture is the probability for the ATI electron 
to return to the core. Intuitively, one could anticipate that 
such a process is suppressed, because the most {\em natural}\/ 
behavior for the electron would be simply to leave the atom. 
The proper description of the suppression plays substantial 
role in the theory. According to the physical image of the ATI 
process worked out in the adiabatic approach \cite{Multa}, 
after tunneling through the time-dependent barrier
the ATI electron emerges from under the barrier at some 
point which is well separated from the core.
As a result this point becomes the source of an expanding 
spherical wave. This occurs twice per each cycle of the 
laser field, at the two moments of time $t_{m \mu}^\prime$ 
when the source-points lie up and down the field ${\bf F}$ 
from the core. The interference of the two spherical waves
originating from the two different source-points results 
in non-trivial patterns in the angular ATI photoelectron
distributions obtained from (\ref{A})-(\ref{Amu})  
\cite{Multa}\cite{Multb} in agreement with 
the available theoretical and experimental data.
The probability for the ATI electron to return to the core  
from the source-point is governed by the expansion factor $1/R$ 
and by the direction of propagation. At each of the moments 
$t_{m \mu}^\prime$ only {\it one}\/ of the two possible 
directions of ${\bf K}_m$, labeled in (\ref{Mpref}) by 
$\sigma = \pm 1$, results in the electron eventually
approaching the core. For the opposite direction of ${\bf K}_m$ 
the electron recedes from the core and does not come back
to recombine. 
In other words, for each direction of ${\bf K}_m$ only one of the
two saddle points $t_{m \mu}^\prime$ contributes to HG.
Since both values of $\sigma$ give identical contributions, 
summation over $\sigma$ simply gives an extra factor of 2 
in (\ref{dfin}).

\subsection{Choice of the gauge}

Calculation of $d_N$ according to the formulae (\ref{dtdef}), 
(\ref{dN}), with exact wave function $\Psi$ might be equivalently 
carried out in various forms which correspond to different choice 
of gauge. As soon as the approximations are employed, the theory 
looses gauge invariance.
Although the length gauge is known to be superior for the description 
of ATI within the adiabatic approximation \cite{Multa}, the situation
is not that straightforward for the high-energy photon.
In the derivation above we employed the dipole-length
form for the operator $\hat{d}_{\epsilon}$ (\ref{dtdef}). 
In the velocity-length form this operator is to be substituted
according to the rule 
($ {\bf p} \equiv - i \nabla_{\bf r}$ is the electron momentum)
\begin{eqnarray} \label{rv}
\hat{d}_{\epsilon} = \, {\bf \epsilon}  \cdot {\bf r} 
\Rightarrow  \, \frac{i}{\Omega} \:{\bf \epsilon}  \cdot {\bf p} 
\end{eqnarray}
when calculation of $d_N$ is concerned ($\Omega = N \omega$).
It is easy to see that this substitution is equivalent to 
replacement of $B_{m \, \mu}$ (\ref{Bs}) by
\begin{eqnarray} \label{Bsv}
B_{m \, \mu}^{(v)} = - \, \frac{1}{\Omega} \, 
\frac{A_a}{\sqrt{\pi} T} \,
\frac{\omega^2}{ F } \,
\int \limits_{0}^{T} dt \:
\frac{\exp \left\{ i \left[\Omega t - S(t) \right] \right\}}
{\cos \omega t - \cos \omega t^\prime_{m \, \mu} } \:
\frac{ {\bf \epsilon} \cdot \left( {\bf K}_m + {\bf k}_t \right)}
{\left[ \left( {\bf K}_m + {\bf k}_t \right)^2 + \kappa^2 \right]} ~.
\end{eqnarray}
As compared with the expression (\ref{Bs}), the latter one differs by 
the extra factor
\begin{eqnarray} \label{ratio}
\frac{\left( {\bf K}_m + {\bf k}_t \right)^2 + \kappa^2 }{2 \Omega}
\end{eqnarray}
in the integrand. 
%This factor could be roughly estimated as
%\begin{eqnarray}
%\frac{\left( {\bf K}_m + {\bf k}_t \right)^2 + \kappa^2 }{2 \Omega}
%\: \sim \: \frac{ {\bf K}_m^2 + \kappa^2 }{2 \Omega}
%\: \sim \: \frac{m \omega - e^2 F^2/(4 \omega^2)}{\Omega} 
%\: \sim \: \frac{m}{N} ~.
%\end{eqnarray}
%Hence results in length and velocity gauges would be close provided
%$d_N$ is mostly contributed by ATI channels with $m \approx N$
%[see (\ref{MGr}), (\ref{Mfin})]. However, moduli of the amplitudes 
%$A_m$ rapidly decrease with the channel number $m$. Therefore on can
%anticipate that the weight of smaller $m$ would be enhanced
%and hence the factor (\ref{ratio}) would be effectively less
%than unity.

\section{Results of calculations} \label{Sec5}

Within framework of the present theory we calculate the rates 
of generating the $N$-th harmonic 
radiation ($c$ is the velocity of light)
\begin{eqnarray} \label{RN}
{\cal R}_N \equiv \frac{\omega^3 N^3}{2 \pi c^3} \, \mid d_N \mid^2 
\end{eqnarray}
introduced by Becker {\it et al}\/ \cite{B} (and denoted by these
authors as $d R_N /d\Omega_{\bf K}$) who in particular carried 
out calculations for the HG by H$^-$ ion in the $\omega = 0.0043$ 
laser field. Fig. \ref{Fig1} provides a complete comparison of 
these results with the results of our calculations.
We employ the binding energy of H$^-$ ($\kappa = 0.2354$), 
but replace the true value 
$A_a=0.75$ \cite{Avalue} by $\sqrt{2 \kappa} = 0.686$ since 
this corresponds to the zero-range potential model 
used by Becker {\it et al}\/ \cite{B}.
Our rates were multiplied by the extra factor $N_e^2$, where
$N_e=2$ accounts for the presence of two active electrons in H$^-$. 
For the real H$^-$ ion the results shown in Fig.\ref{Fig1} are
to be scaled by a factor $A_a^4 / (2 \kappa)^2$.

Fig. \ref{Fig1} shows harmonic spectra for the laser field
intensities $I = 10^{10}, \: \, 2 \cdot 10^{10}, \: \,  
5 \cdot 10^{10}, \: \, 10^{11}$ W/cm$^2$ that correspond respectively
to the Keldysh adiabatic parameter values 
$\gamma \equiv \omega \kappa / F = 1.898, \: \, 1.342, \: \, 
0.849, \: \, 0.600$. Thus the most interesting region
of transition from
the multiphoton regime ($\gamma \gg 1$) to the
tunneling mechanism ($\gamma \ll 1$) is covered. 
Our major results are shown by open circles in Fig. \ref{Fig1}.
They are obtained using the expression (\ref{dfin}) for $d_N$
with the time-integration in $B_{N \, m \mu_0}$ (\ref{Bs}) 
carried out numerically. 
Based on physical arguments, we extend the summation 
in (\ref{dfin}) only over open ATI channels with the 
real values of $K_m$.
Generally the HG spectrum is known to consist of the initial
rapid decrease, the plateau domain and the rapid fall-off region.
The present theory is designed to describe the high
HG but not the initial decrease, which in the case considered 
is noticeable only for one or two lowest harmonics.
In the fall-of region, i.e.
on the large-$N$ side, our rates perfectly coincide with those 
obtained by Becker {\it et al}\/ \cite{B} (closed circles in 
Fig. \ref{Fig1}). The slight difference that could be hardly
distinguished in the plot scale lies within uncertainty
%!   below
in retrieving data from the small-size plot published in Ref. \cite{B}.
Of course, our comparison is carried out in absolute scale
without any fitting or normalization.
%!   above

The deviations increase as $N$ decreases, but within entire
plateau region the agreement of rates averaged over 
structures remains good. 
%!   below
Remarkably, the positions of numerous dips and peaks 
that exist in the plateau region are well reproduced
by our calculations, although there exist some, 
generally not strong, discrepancies in their magnitudes. 
The structures has not yet received a universally accepted 
physical interpretation in the current literature with 
two tentative explanations being available. 
Becker {\it et al}\/ \cite{Brapid}
relate these structures to ATI channel closing
whereas Lewenstein {\it et al}\/ \cite{phase} suggest that
it stems from quantum interferences between the contribution
of different electron trajectories.
%!    above 
Within the present framework we can say that the origin 
of these structures lies in the interplay of interfering 
contributions of various ATI paths, but their precise description 
would require an additional detailed analysis. 

Approximation (\ref{Ma}) with numerical 
calculation of the integral in $C_{N m}$ (\ref{CF}) (open squares in 
Fig. \ref{Fig1}) somewhat overestimates HG rate, but still 
retains the structure, though smoothed.
It is worthwhile to emphasize two circumstances. First, to the best 
%!    below
of our knowledge, we present here almost unique quantitative 
comparison of HG rates calculated within different theoretical 
methods, namely, by the present approach and by that developed by
Becker {\it et al}. 
We are aware of only two cases when the quantitative comparison 
was carried out previously, one refers to some parameters
%!!
of HG by elliptically polarized light in Ref.~\cite{Buni}
(see also discussion in the Introduction), the other is 
comparison between the approach of Ref.~\cite{Lew}
and results of direct numerical integration of the
three-dimensional time-dependent Schr\"{o}dinger approximation
\cite{Tempo}.
%!    above
Second, the comparison is presented in
the log-scale, as are the results reported by Becker {\it et al},
because it is appropriate both for the physics of the problem
and for the current state of experiment. 

In the summation (\ref{dfin}) over ATI channels (i.e. over $m$) 
the coherence is very important, since large number of terms 
is comparable in modulus, but have rapidly varying phases.
Many ATI channels contribute to HG for each $N$.
This is in variance with the tentative conclusion 
by Eberly {\it et al}\/ \cite{Eb}).
The low ATI channels give appreciable contribution even
for quite high harmonics. Only for the highest harmonics 
considered the contribution of low ATI channels becomes 
negligible.

Although the length gauge is known to be superior for the description 
of ATI within the adiabatic approximation \cite{Multa}, the situation
is not that straightforward for the high-energy photon.
Therefore our calculations for the rates were carried out 
using both the length and velocity gauges.  
For large $N$ the results obtained are very close, see Fig. \ref{Figratio};
the dipole-velocity gauge producing the rates about 10\% larger.
For smaller $N$ the difference increases and manifests rather
irregular $N$-dependence. However, even in the most unfavorable
situation the ratio of the length-form to velocity-form results
deviates from 1 not more than by 50\%, which is rather
reasonable, bearing in mind the multiphoton nature of the process.

An evolution of the parameters of the individual $N$-th harmonic 
radiation with variation of the laser field intensity $I$ 
is presented in Fig. \ref{Figphase} for some particular values of $N$.
We show both $\left| d^+_N \right|^2$ (that is proportional 
to HG rate) and the phase $\Phi_N \equiv \arg d^+_N$.
The intensity dependence of the phase $\Phi_N$ is known 
\cite{Zar2,phaselet,phase,CohCont,temporal}
to play an important role in description of the harmonic field 
propagation in the experimental conditions when the spectral 
and spatial coherence properties are substantial. 
Our calculations demonstrate rapid variation of the phase 
$\Phi_N$ with the intensity $I$: as $I$ increases by an order 
of magnitude the phase changes by about $10 \pi$ (for $N=9$)
or $15 \pi$ (for $N=11$). The dependence of phase on $I$ looks 
most simple in the fall-off domain where it is essentially
linear, Fig. \ref{Figphase}d.
As it is well known \cite{Brapid,B}, the HG rate manifests 
spikes (see Fig. \ref{Figphase}a, where the spikes are slightly
smoothed due to finite step over $I$ used in plotting) 
at the intensities $I_m$ that correspond to the threshold of 
$m$-th ATI channels, that is closed for
$I > I_m$ due to ponderomotive potential $F^2/( 4 \omega^2)$).

As discussed in Sec.\ref{Sec2}, in the Fourier component 
$d_N$ (\ref{dnmn}) the contribution of $d^+_{-N}=d^-_N$ 
is anticipated to be negligible as compared with that of 
$d^+_N$. The reader have to recall that the physical reason 
for this is that $d^+_N$ describes the {\it natural}\/ 
sequence of events, when electron at first absorbs energy 
from the laser field and subsequently emits the high harmonic
photon, whereas for $d^-_N$ the sequence is inverted.
In order to illustrate how strong the preference is we 
show in Fig. \ref{Figminus} the ratio
$\left| d^-_N / d^+_N \right|^2$. The ratio rapidly
decreases with $N$ being very small in all cases
where the present theory applies. It becomes noticeable
only for small $N$ in case of intense laser field.

\section{CONCLUSION} \label{Sec6}

The major result of this paper is a quantum mechanical
description of the high-harmonic generation problem as 
a three-step process. In the first step the atomic electron 
absorbs several laser quanta and populates some ATI channel. 
Secondly, the electron propagates in the laser field back 
to the atomic core. The third step is the laser-induced
recombination when the high harmonic photon is emitted. 
This mechanism is nothing more but interpretation
of our principal formulae (\ref{dfin}), (\ref{Ma})
derived quantum mechanically with minimal approximations
and without resort to any classical 
%!   below
or intuitive arguments.
The distinctive feature of our formulae (\ref{dfin}), (\ref{Ma})
is that they include only genuine amplitudes of the three
constituent processes, i.e. each amplitude describes
true physical fully accomplished process, as HG, ATI or LAR.
Here lies a crucial difference between our results and
those of previous authors \cite{Lew} (see also Introduction) 
who discussed mathematical structure of the matrix elements 
or integrals considered in their approach in terms of three-step 
mechanism but failed to present HG amplitude via amplitudes of 
physically observable accomplished ATI and LAR processes.
%!   above
Even in a more broad context 
this is a rare and remarkable situation when a complicated process 
is reduced essentially exactly to the sum over all possible 
paths with every path described as a sequence of real, fully 
accomplished physical processes. This conceptual  
simplicity arises due to multiphoton, adiabatic nature of HG process.

With all three constituent amplitudes available from analytical
formulae or simple numerical calculations our theory
promises to be an efficient practical tool. This hope is 
supported by good agreement of our quantitative
results for HG by H$^-$ ion with the previous calculations by
Becker {\it et al}\/ \cite{B} in a wide range of laser 
intensities and frequencies of the emitted quantum.
As the simplest example of possible future extensions we 
note only that our approach can be straightforwardly applied 
to negative ions with the outer electron having
non-zero orbital momentum, such as halogen ions. 
These species could be easier accessible for the experimental 
studies. 

%!   below
The structure of the formulae (\ref{dfin}), (\ref{Ma}) is 
so simple that it is tempting to suggest that they will
work well if one substitutes in them ATI or LAR amplitudes
obtained in some approximations other than those used in
the derivation above. For instance, if ATI amplitude
more accurate than that given by the Keldysh approximation
is available from some theory, one can employ it for
HG calculations via (\ref{dfin}), (\ref{Ma}). This gives
a hope to relatively simply improve account 
for the electron rescattering effects in the HG theory.
%!   above

The distinctive feature of the present theory of HG is the use 
of the representation based on the discrete set of ATI channels.
To the best of our knowledge this simple and apparently 
straightforward idea has not been exploited before. It would be
underestimate to consider it merely as a detail of theoretical
technique. Indeed, if one pursues the objective of the
most direct and far-reaching quantum implementation of
the three-step mechanism of HG, then the use of ATI
channel representation becomes an unavoidable and crucial point. 
Otherwise, if the amplitudes of ATI do not appear in 
the theoretical scheme, HG cannot be properly described 
as a three-step process.

In this paper we do not discuss qualitative features
of HG spectra, such as extension of plateau domain etc.
Detailed discussion of these issues could be found in
other publications \cite{C,Hu,Lew}, in 
particular in the paper by Becker {\it et al}\/ \cite{B} 
whose results for HG by H$^-$ ion we reproduce closely 
within our theory. This implies that the analysis of the
{\it numerical results}\/ carried out by Becker 
{\it et al}\/ \cite{B} is fully applicable in our case. 
Concerning the {\it mechanisms and physical interpretation}\/
of HG spectra, the present theory, hopefully, can add more 
to our understanding. However, the related analysis requires 
further theoretical developments which would overburden 
the present, already quite a long paper. We hope to present 
these subsequent developments elsewhere. 
In a broader perspective, one can expect that modifications of 
our approach  could be applied to a variety of processes
such as population of high ATI channels and multiple 
ionization of atoms.

As a summary, the three-step mechanism of the harmonic
generation is ultimately justified. 
% It is implemented in fully	
% quantum relations expressing its amplitude via amplitudes
% of the above-threshold ionization and stimulated recombination.
The theory is quantitatively reliable and easy to apply.
It gives an important physical insight being a particular
realization of the general {\it atomic antenna}\/ mechanism.

\acknowledgements

This work has been supported by the Australian Research Council. 
V.~N.~O. acknowledges the hospitality of the staff of 
the School of Physics of UNSW where this work has been
carried out.

\begin{figure}
\caption{ \label{Fig1}
Harmonic generation rates (\protect \ref{RN})
(in sec$^{-1}$) for H$^-$ ion 
in the laser field with the frequency $\omega = 0.0043$ 
and various values of intensity $I$ as indicated in the plots.
Closed circles - results obtained by Becker {\it et al}\/
\protect\cite{B}, open circles - present calculations
in the dipole-length gauge
using the expression (\protect \ref{Bs}) for $B_{N m \, \mu}$,
open squares - same but with the simplified formula
(\protect \ref{Ma}) for $B_{N m \, \mu}$.
}
\end{figure}

\begin{figure}
\caption{\label{Figratio}
Ratio of rates for HG by H$^-$ ion in the laser field with 
the frequency $\omega = 0.0043$ calculated within the present 
approach using the velocity and length gauges. 
(a) -- for the rates calculated using the expression 
(\protect \ref{Bs}) for $B_{N m \, \mu}$:
circles -- $I = 10^{10}$; 
squares -- $I = 2 \cdot 10^{10}$;
triangles -- $I = 5 \cdot 10^{10}$;
diamonds -- $I =  10^{11}$ W/cm$^2$;
(b) - same as in (a) but with the simplified formula
(\protect \ref{Ma}) for $B_{N m \, \mu}$.
}
\end{figure}

\begin{figure}
\caption{\label{Figphase}
Evolution of HG parameters with variation of the laser 
field intensity $I$. The harmonic intensity parameter 
$M_N \equiv 2 \log_{10} \left| d^+_N \right|$ and 
the reduced harmonic phase $\Phi_N/\pi$ with
$\Phi_N \equiv \arg d^+_N$ are shown for harmonic 
of various order $N$.
The bars with numbers $m$ indicate the threshold intensities
$I_m$ such that for $I>I_m$ the $m$-th ATI channel is closed
due to ponderomotive potential. 
}
\end{figure}

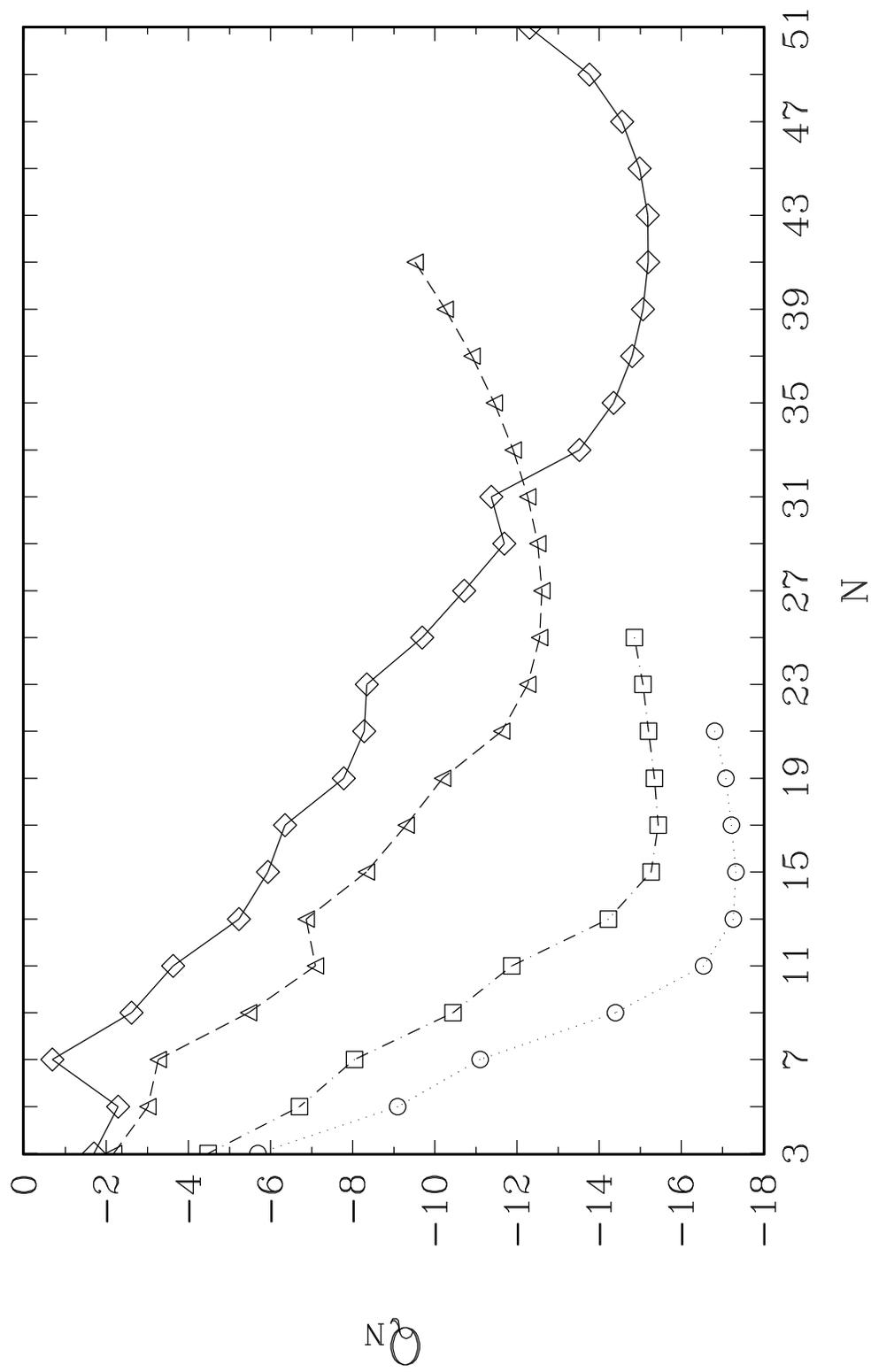
\begin{figure}
\caption{\label{Figminus}
Parameter $Q_N = 2 \log_{10} \left| d^-_N / d^+_N \right|$
as a function of harmonic number $N$
for HG by H$^-$ ion in the laser field with the frequency
$\omega = 0.0043$ and various intensities: 
circles -- $I = 10^{10}$; 
squares -- $I = 2 \cdot 10^{10}$;
triangles -- $I = 5 \cdot 10^{10}$;
diamonds -- $I =  10^{11}$ W/cm$^2$.
}
\end{figure}

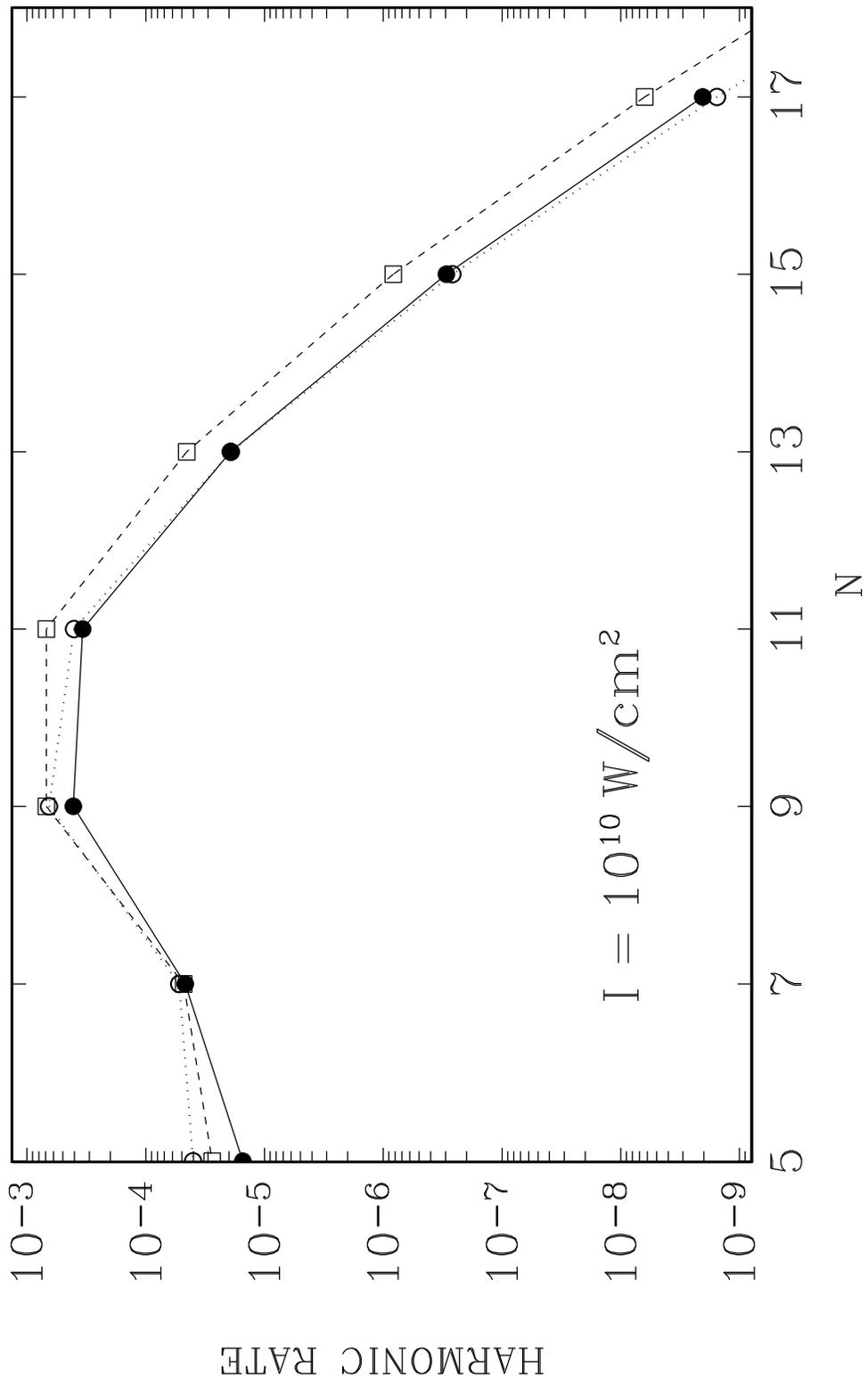
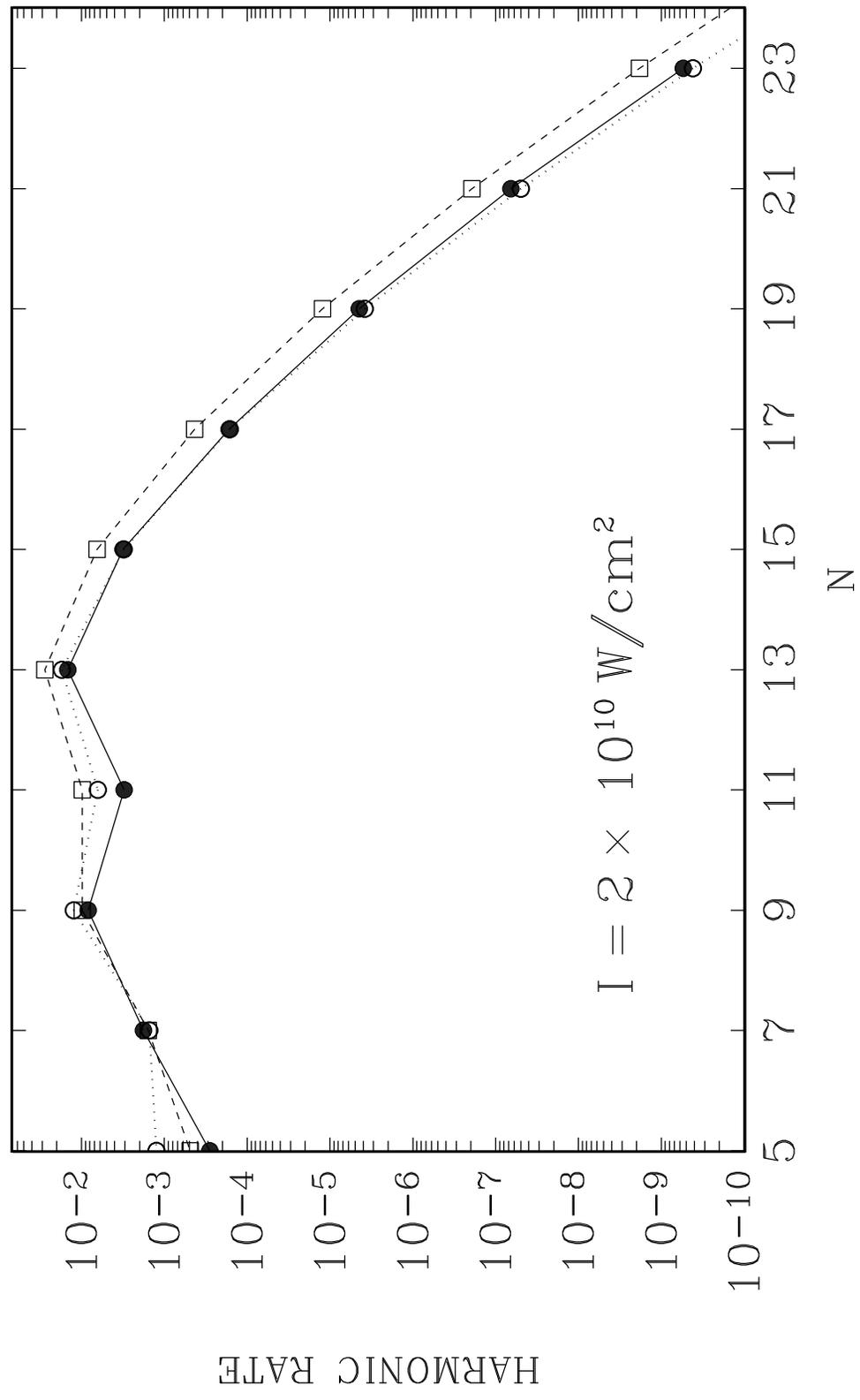
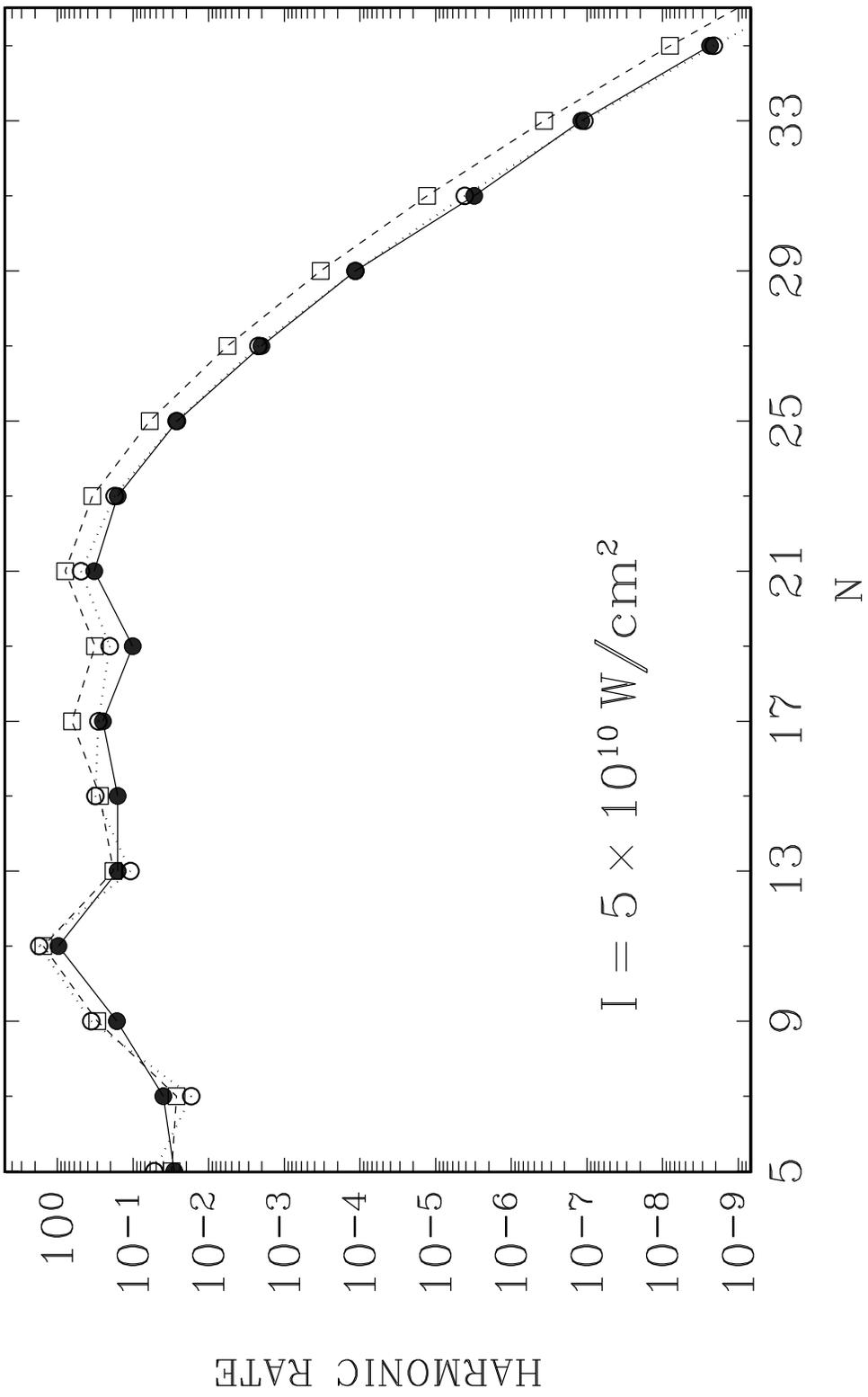
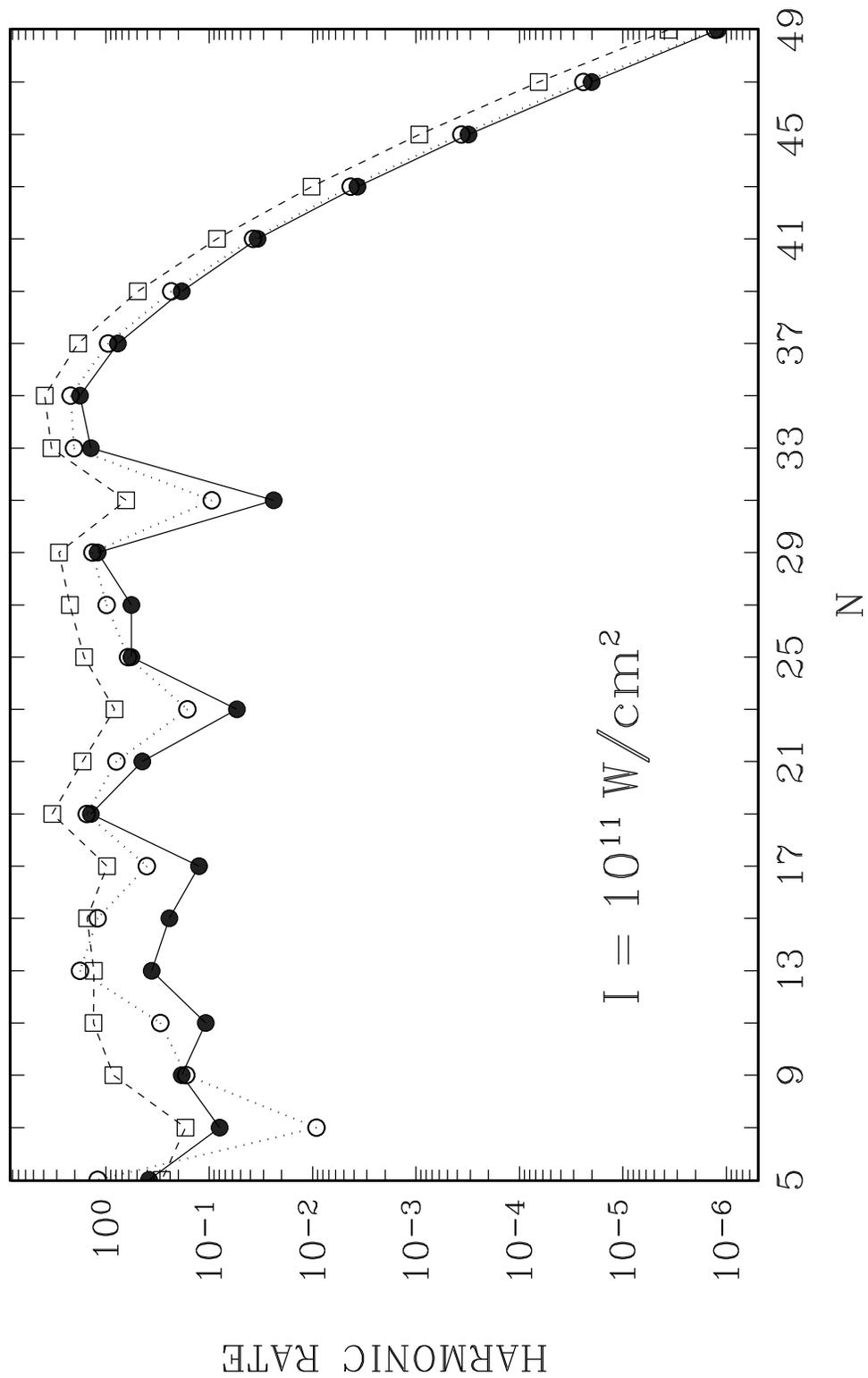
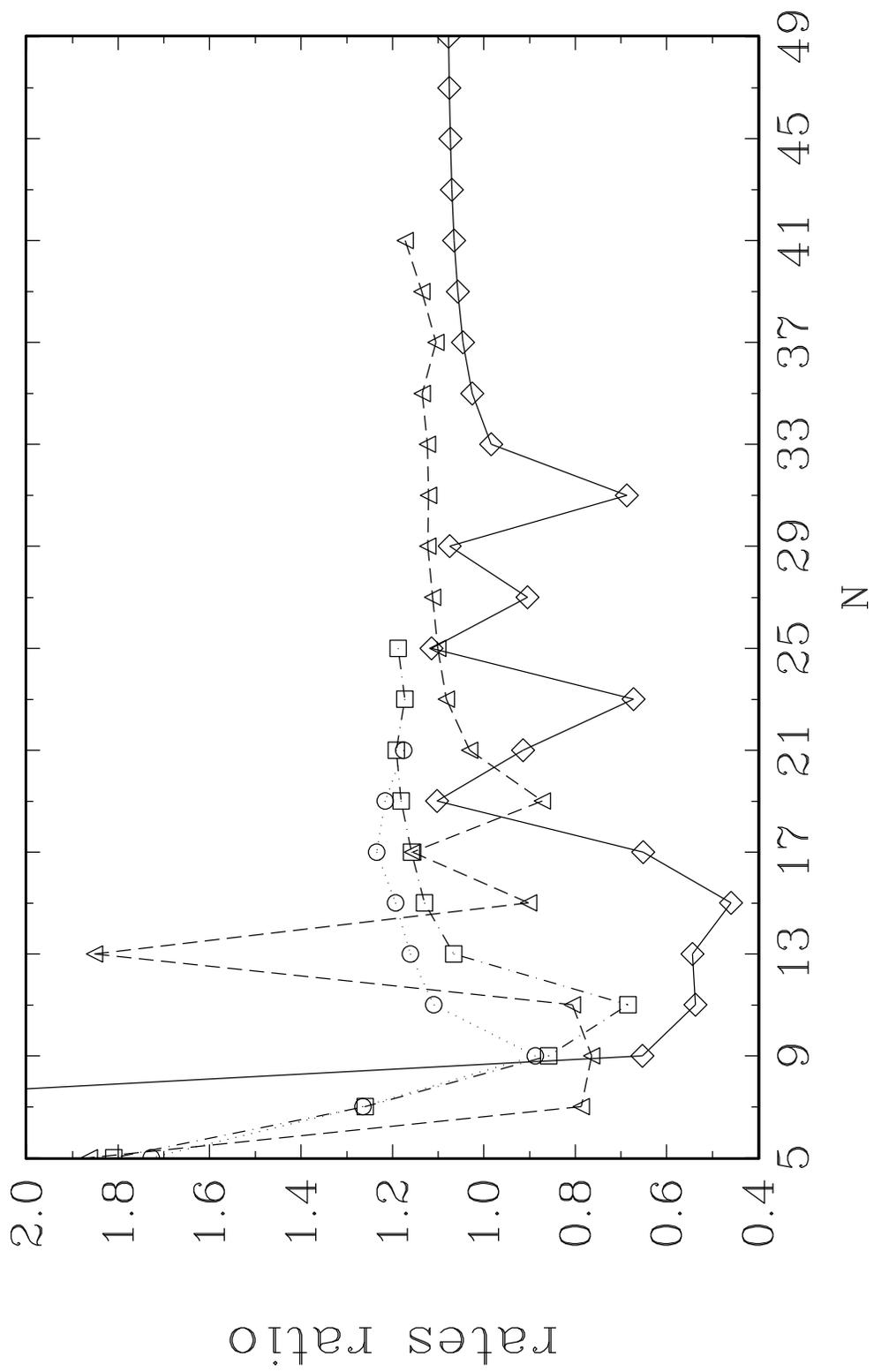
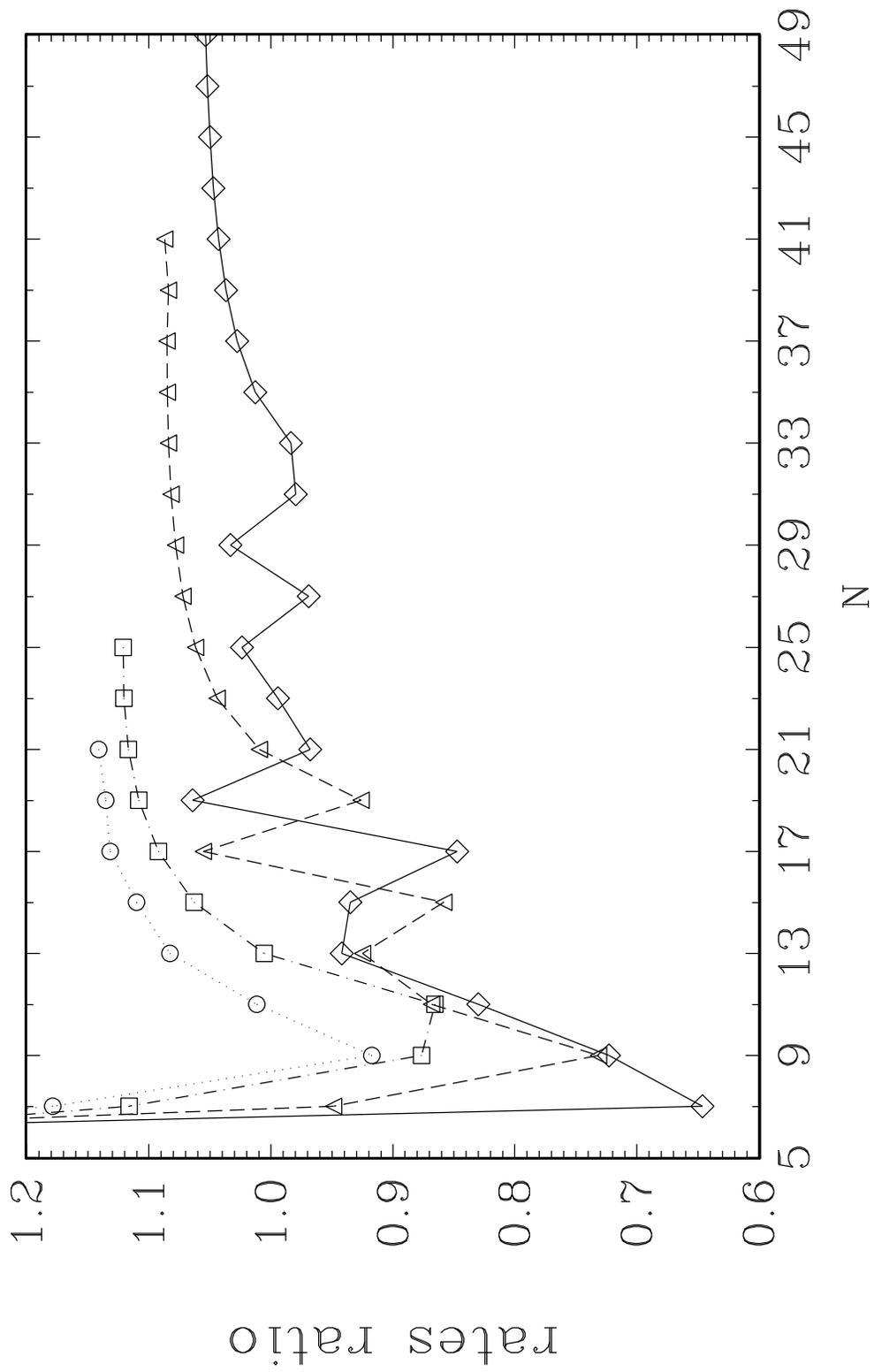
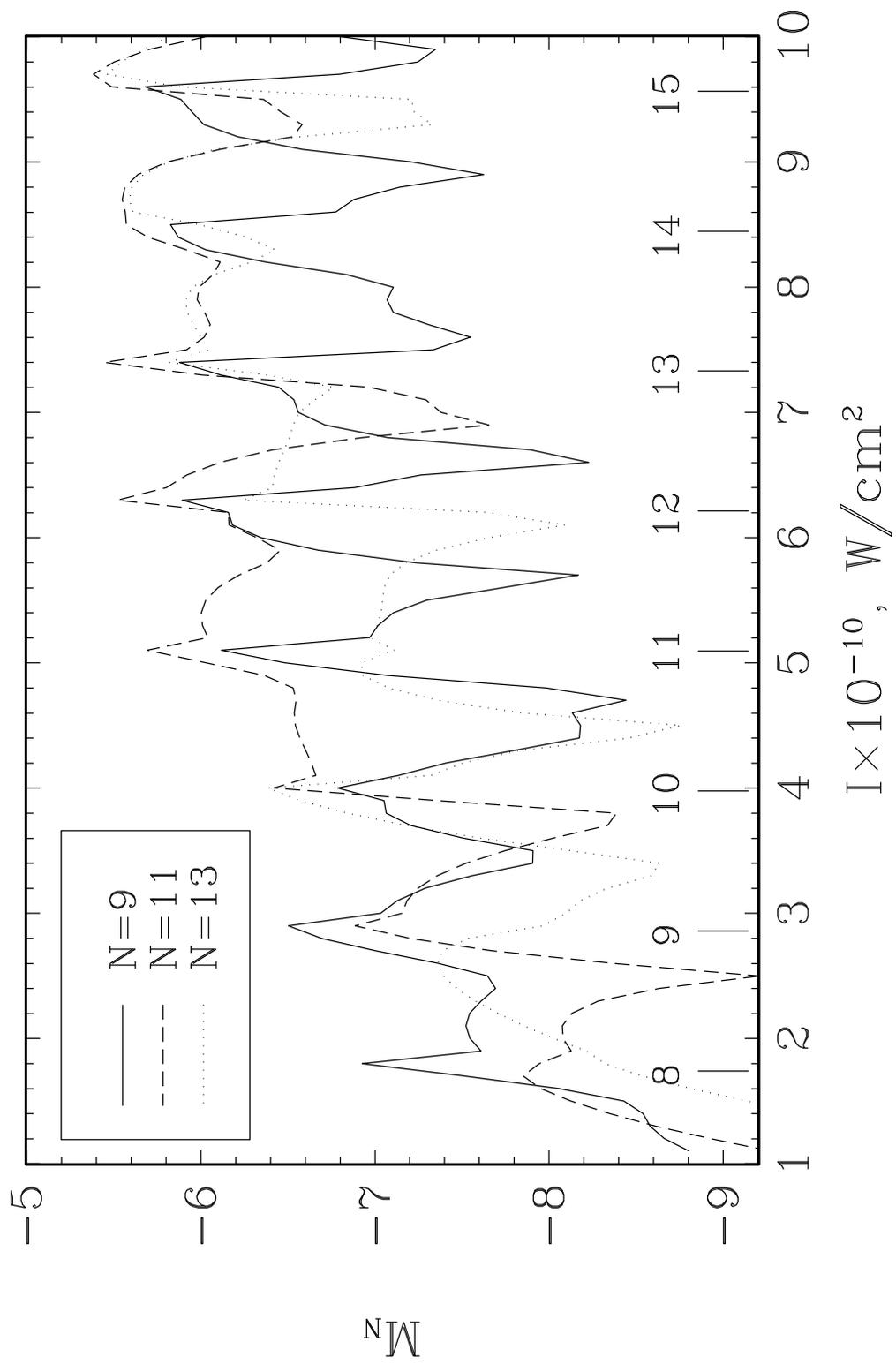
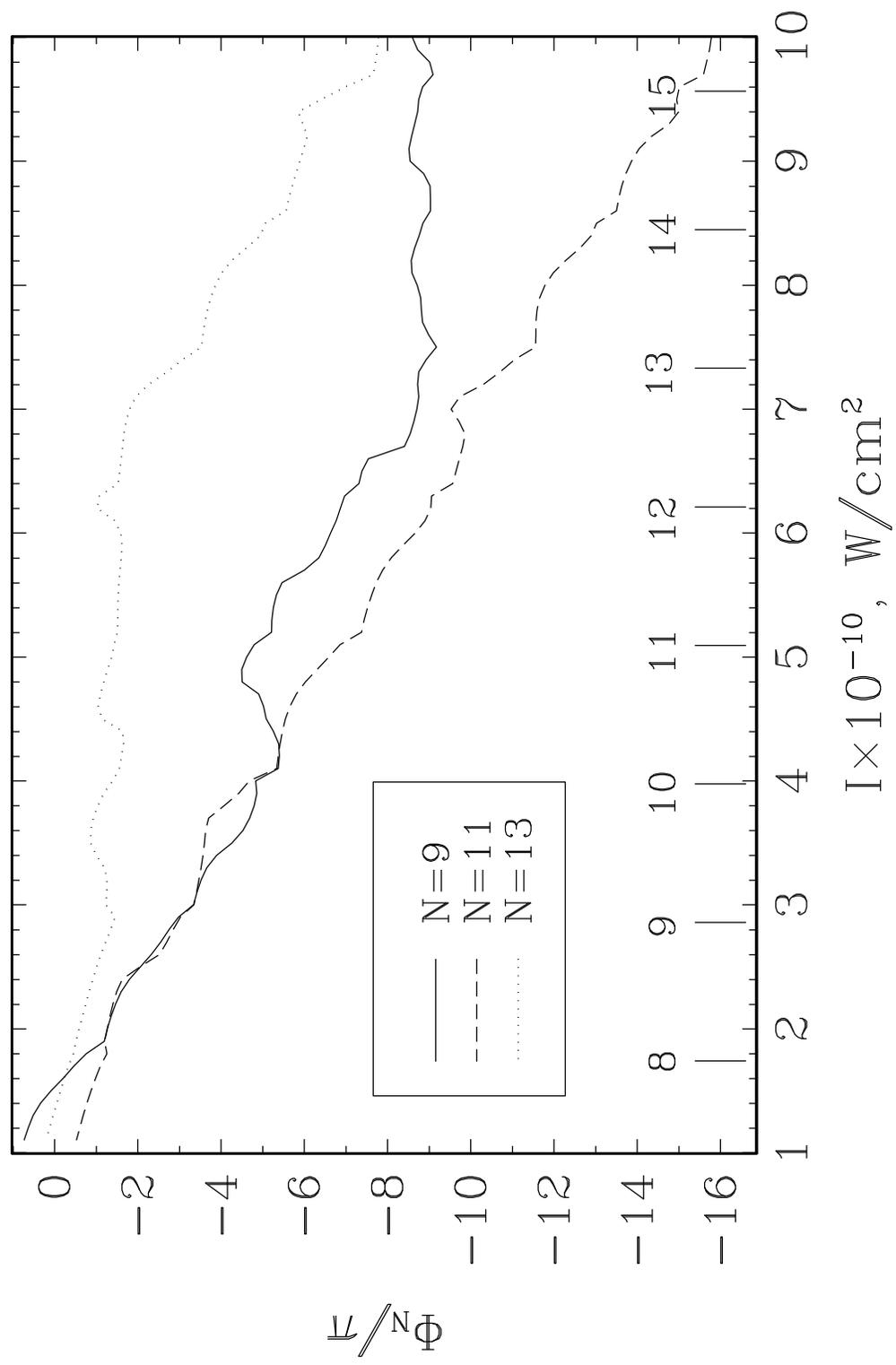
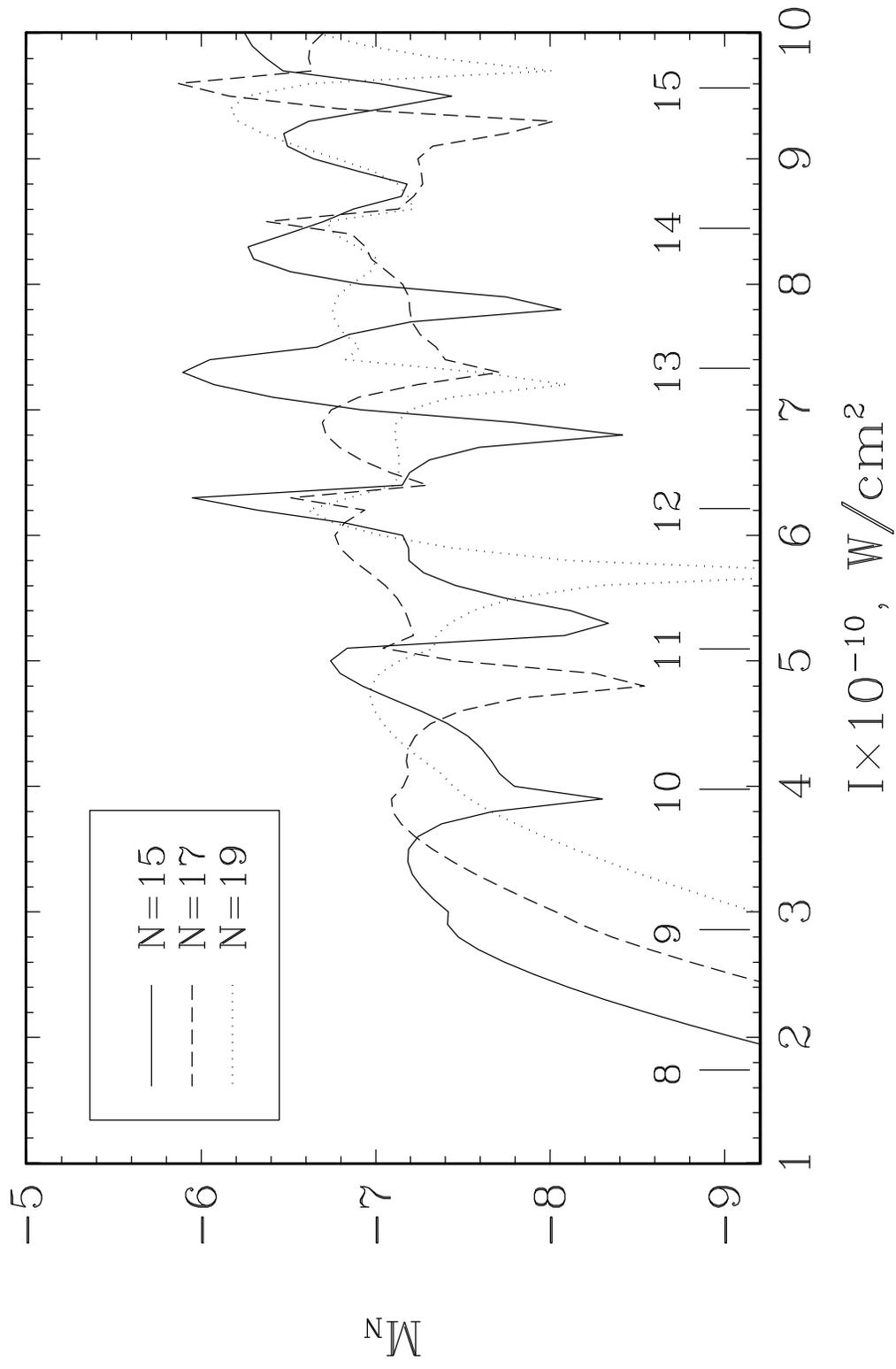
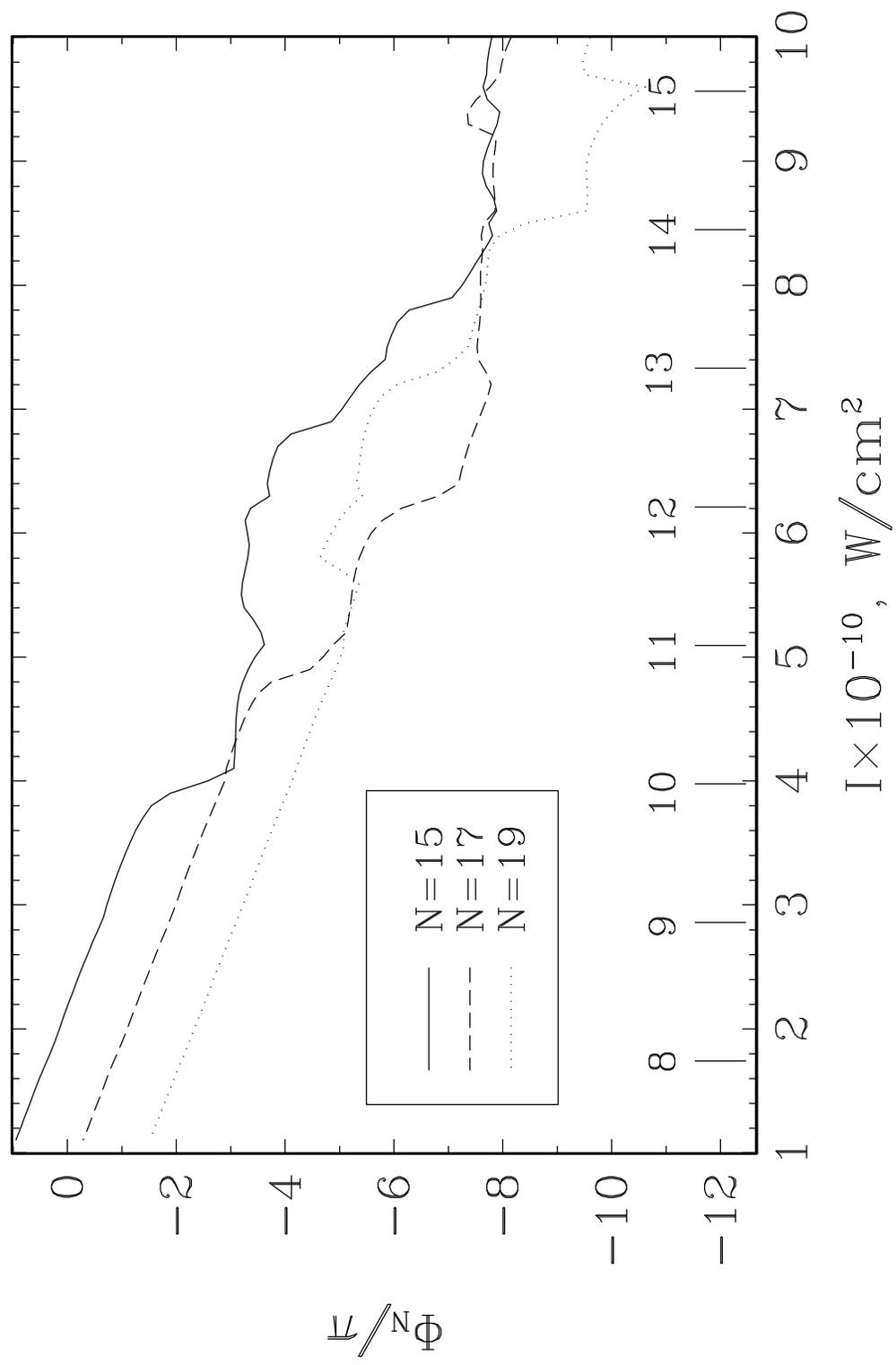
\end{document}